%% file: kv-match.extended.tex
\documentclass[10pt,conference,letterpaper]{IEEEtran}

\usepackage{graphicx}
\usepackage{balance}
\usepackage{algorithm}
\usepackage[noend]{algpseudocode}
\usepackage{tabulary}
\usepackage{amsmath,bm}
\usepackage{amsthm}
\usepackage{mathtools}
\usepackage{calc}
\usepackage{amsfonts,amssymb}
\usepackage{url}
\usepackage{multirow}
\usepackage{diagbox}
\usepackage{enumerate}
\usepackage{needspace}
\usepackage[perpage]{footmisc}
\usepackage{xcolor}
\usepackage{textcase}
\usepackage{capt-of}

\usepackage{paralist}

\newtheorem{definition}{Definition}
\newtheorem{lemma}{Lemma}
\newtheorem{property}{Property}

\def\comment#1{}

\newenvironment{lcases}
{\left\lbrace\begin{aligned}}
	{\end{aligned}\right.}

\title{KV-match: A Subsequence Matching Approach Supporting Normalization and Time Warping\\ \large{[Extended Version]}}

\author{
{Jiaye Wu{\small $~^{\#}$}, Peng Wang{\small $~^{\#}$}, Ningting Pan{\small $~^{\#}$}, Chen Wang{\small $~^{*}$}, Wei Wang{\small $~^{\#}$}, Jianmin Wang{\small $~^{*}$} }
\vspace{1.6mm}\\
\fontsize{10}{10}\selectfont\itshape
$^{\#}$\,School of Computer Science, Fudan University, Shanghai, China\\
\fontsize{9}{9}\selectfont\ttfamily\upshape
\{wujy16, pengwang5, ntpan17, weiwang1\}@fudan.edu.cn
\vspace{1.2mm}\\
\fontsize{10}{10}\selectfont\rmfamily\itshape
$^{*}$\,School of Software, Tsinghua University, Beijing, China\\
\fontsize{9}{9}\selectfont\ttfamily\upshape
\{wang\_chen, jimwang\}@tsinghua.edu.cn
}

\begin{document}
\maketitle

\begin{abstract}
	The volume of time series data has exploded due to the popularity of new applications, such as data center management and IoT. Subsequence matching is a fundamental task in mining time series data. All index-based approaches only consider raw subsequence matching (RSM) and do not support subsequence normalization. UCR Suite can deal with normalized subsequence matching problem (NSM), but it needs to scan full time series. In this paper, we propose a novel problem, named constrained normalized subsequence matching problem (cNSM), which adds some constraints to NSM problem. The cNSM problem provides a knob to flexibly control the degree of offset shifting and amplitude scaling, which enables users to build the index to process the query. We propose a new index structure, KV-index, and the matching algorithm, KV-match. With a single index, our approach can support both RSM and cNSM problems under either ED or DTW distance. KV-index is a key-value structure, which can be easily implemented on local files or HBase tables. To support the query of arbitrary lengths, we extend KV-match to KV-match$_{\text{DP}}$, which utilizes multiple varied-length indexes to process the query. We conduct extensive experiments on synthetic and real-world datasets. The results verify the effectiveness and efficiency of our approach.
\end{abstract}

\input{tex/introduction1.extended}

\input{tex/preliminary1}

\input{tex/theoretical.extended}

\input{tex/index.extended}
\input{tex/matching}

\input{tex/dynamic1.extended}
\input{tex/implementation}
\input{tex/experiment.extended}
\input{tex/related1}
\input{tex/conclusion}

\bibliographystyle{IEEEtran}

\enlargethispage{8mm}

\input{tex/appendices}

\end{document}

%% file: tex/introduction1.extended.tex
\section{Introduction}
\label{sec:introduction}
Time series data are pervasive across almost all human endeavors, including medicine, finance and science. In consequence, there is an enormous interest in querying and mining time series data.~\cite{fast_shapelets, matrix_profile}. 

Subsequence matching problem is a core subroutine for many time series mining algorithms.
Specifically, given a long time series $X$, for any query series $Q$ and a distance threshold $\varepsilon$, the subsequence matching problem finds all subsequences from $X$, whose distance with $Q$ falls within the threshold $\varepsilon$. 


FRM~\cite{frm94} is the pioneer work of subsequence matching. Many approaches have been proposed, either to improve the efficiency~\cite{vldb12,gmatch02} or to deal with various distance functions~\cite{tods11,vldb07}, such as Euclidean distance and Dynamic Time Warping. However, \emph{all} these approaches only consider the raw subsequence matching problem (RSM for short).
In recent years, researchers realize the importance of the subsequence normalization~\cite{kdd12}. It is more meaningful to compare the z-normalized subsequences, instead of the raw ones. UCR Suite~\cite{kdd12} is the state-of-the-art approach to solve the normalized subsequence matching problem (NSM for short).

The NSM approach suffers from two drawbacks. First, it needs to scan the full time series $X$, which is prohibitively expensive for long time series. For example, for a time series of length $10^9$, UCR Suite needs more than 100 seconds to process a query of length 1,000. \cite{kdd12} analyzed the reason why it is impossible to build the index for the NSM problem. Second, the NSM query may output some results not satisfying users' intent. The reason is that NSM fully ignores the offset shifting and amplitude scaling. However, in real world applications, the extent of offset shifting and amplitude scaling may represent certain specific physical mechanism or state. Users often only hope to find subsequences within similar state as the query. We illustrate it with an example.

\linespread{0.98}

\begin{figure}
	\centering
	\footnotesize
	\begin{tabular}{@{}crr@{}}
		\multirow{8}{0.6\linewidth}{\includegraphics[width=1.1\linewidth]{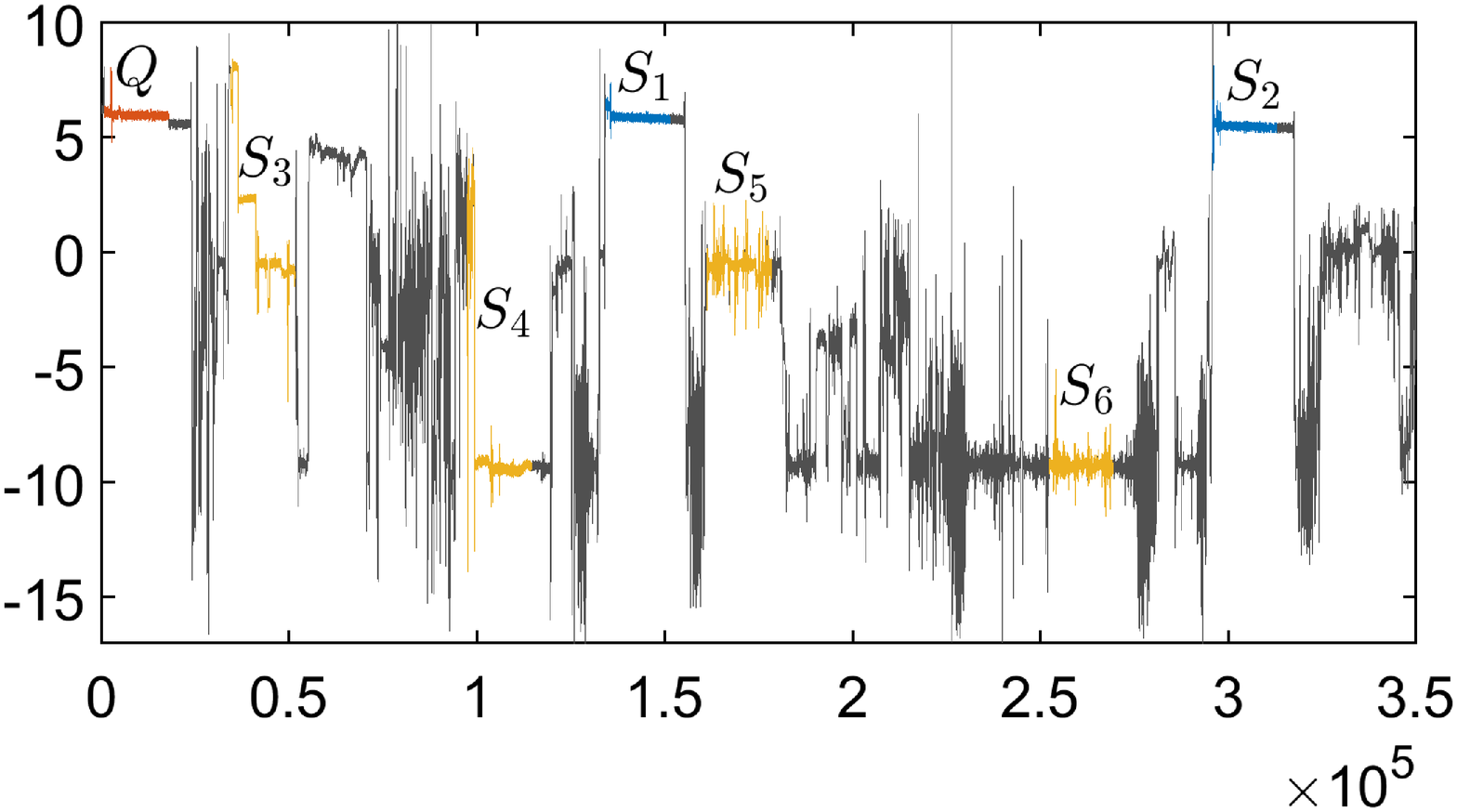}}~
		& & \\
		& \multicolumn{2}{c}{(c) Information of Q}\\[1mm]
		\cline{2-3}
		& Offset \scriptsize{[Label]} & Length \\
		\cline{2-3}
		& 877 \scriptsize{[$Q$]} & 17,124 \\
		\cline{2-3}
		& & \\
		& \multicolumn{2}{c}{(d) Results of NSM}\\[1mm]
		\cline{2-3}
		& Offset \scriptsize{[Label]} & Distance \\
		\cline{2-3}
		& 252,492 \scriptsize{[$S_6$]} & 117.78 \\
		(a) PAMAP time series & 97,458 \scriptsize{[$S_4$]} & 130.80 \\
		\multirow{6}{0.6\linewidth}{\includegraphics[width=1.1\linewidth]{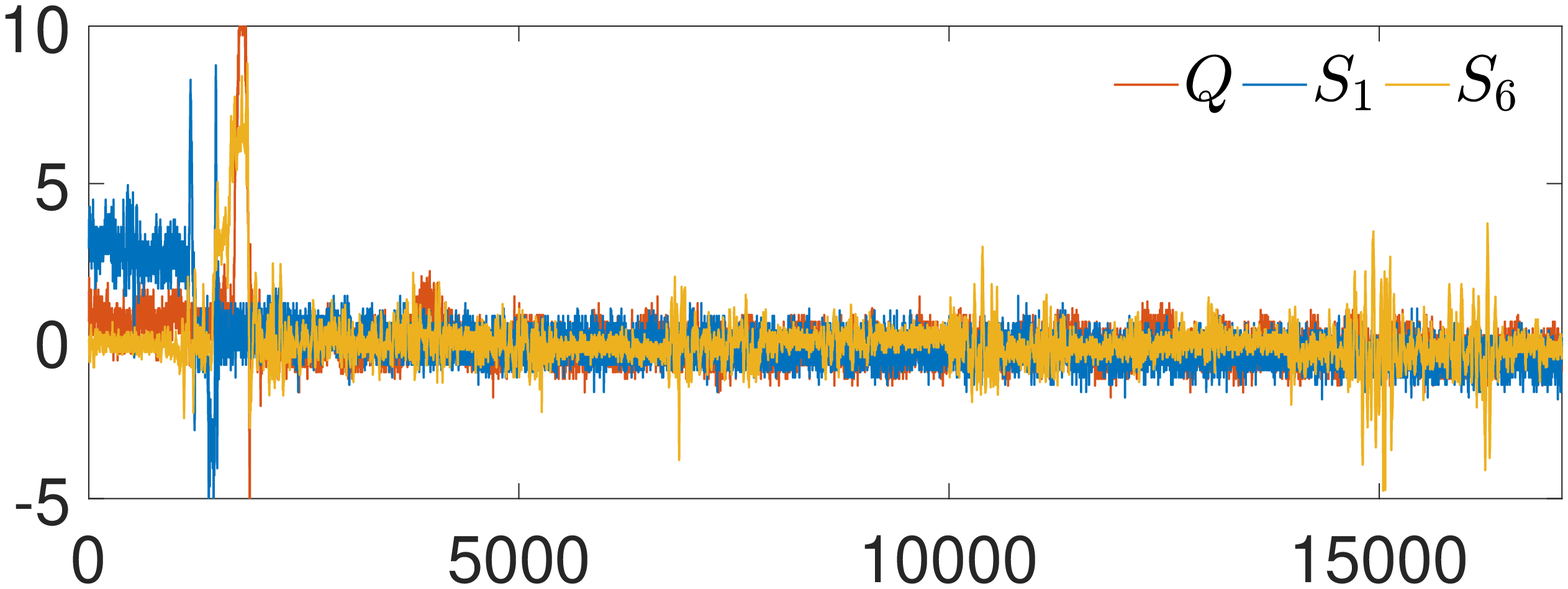}}~ & 34,562 \scriptsize{[$S_3$]} & 138.12 \\
		& 161,416 \scriptsize{[$S_5$]} & 149.37 \\
		& $\cdots$ & $\cdots$ \\
		& 134,456 \scriptsize{[$S_1$]} & 164.88 \\ 
		& 296,063 \scriptsize{[$S_2$]} & 166.74 \\
		& $\cdots$ & $\cdots$ \\
		\cline{2-3}
		(b) Aligned normalized subsequences & \multicolumn{2}{l}{$\star~~ \varepsilon=200.0$} \\
	\end{tabular}
	\caption{\label{fig:cnsm2}Illustrative example of cNSM}
\end{figure}

\linespread{0.91}

\textbf{Example 1.} The time series in Fig.~\ref{fig:cnsm2}(a) comes from the Physical Activity Monitoring for Aging People (PAMAP) dataset~\cite{fast_shapelets} collected from z-accelerometer at hand position. The monitored person conducts various activities alternatively, like sitting, standing, running and so on. Each activity lasts for about 3 minutes, and the data collection frequency is 100Hz. We use one subsequence corresponding to lying activity as the query ($Q$ in Fig.~\ref{fig:cnsm2}(c)) to find other ``lying'' subsequences. We issue a NSM query with $Q$, and Fig.~\ref{fig:cnsm2}(d) lists the top results. Unfortunately, all top-4 results corresponds to other activities. $S_3$ and $S_5$ correspond to sitting activity, while $S_4$ and $S_6$ correspond to breaking activity. Although $S_1$ and $S_2$ are the desired results (correspond to lying activity), they are ranked out of top-20. We show the normalized $Q$, $S_1$ and $S_6$ in Fig.~\ref{fig:cnsm2}(b). It is difficult to distinguish them after normalization.

\begin{figure}[t]
	\begin{minipage}[t]{0.55\linewidth}
		\centering
		\raisebox{0.2\height}{\includegraphics[width=0.9\linewidth]{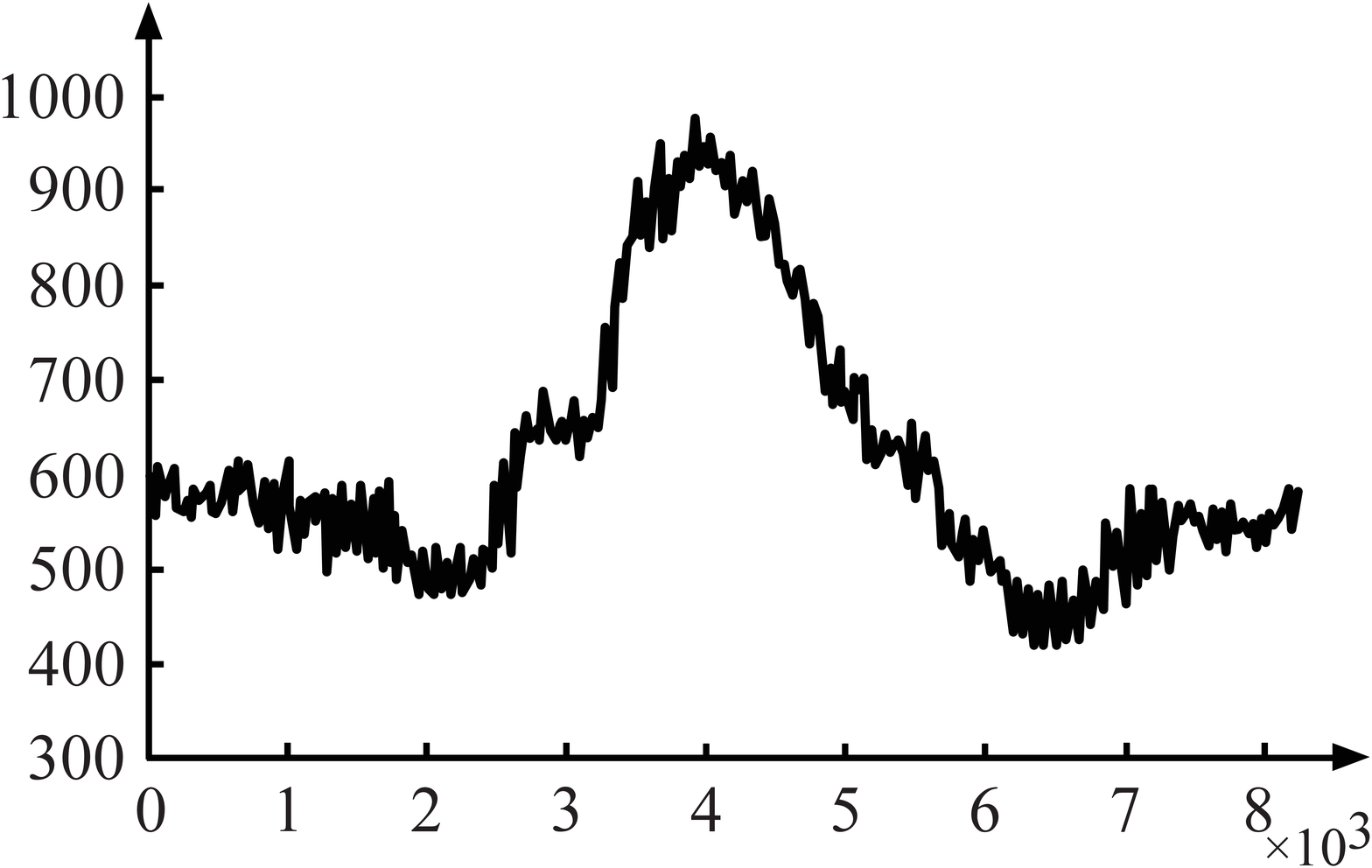}}
		\caption{\label{fig:wind}EOG pattern}
	\end{minipage}%
	\hfill%
	\begin{minipage}[t]{0.44\linewidth}
		\centering
		\includegraphics[width=\linewidth]{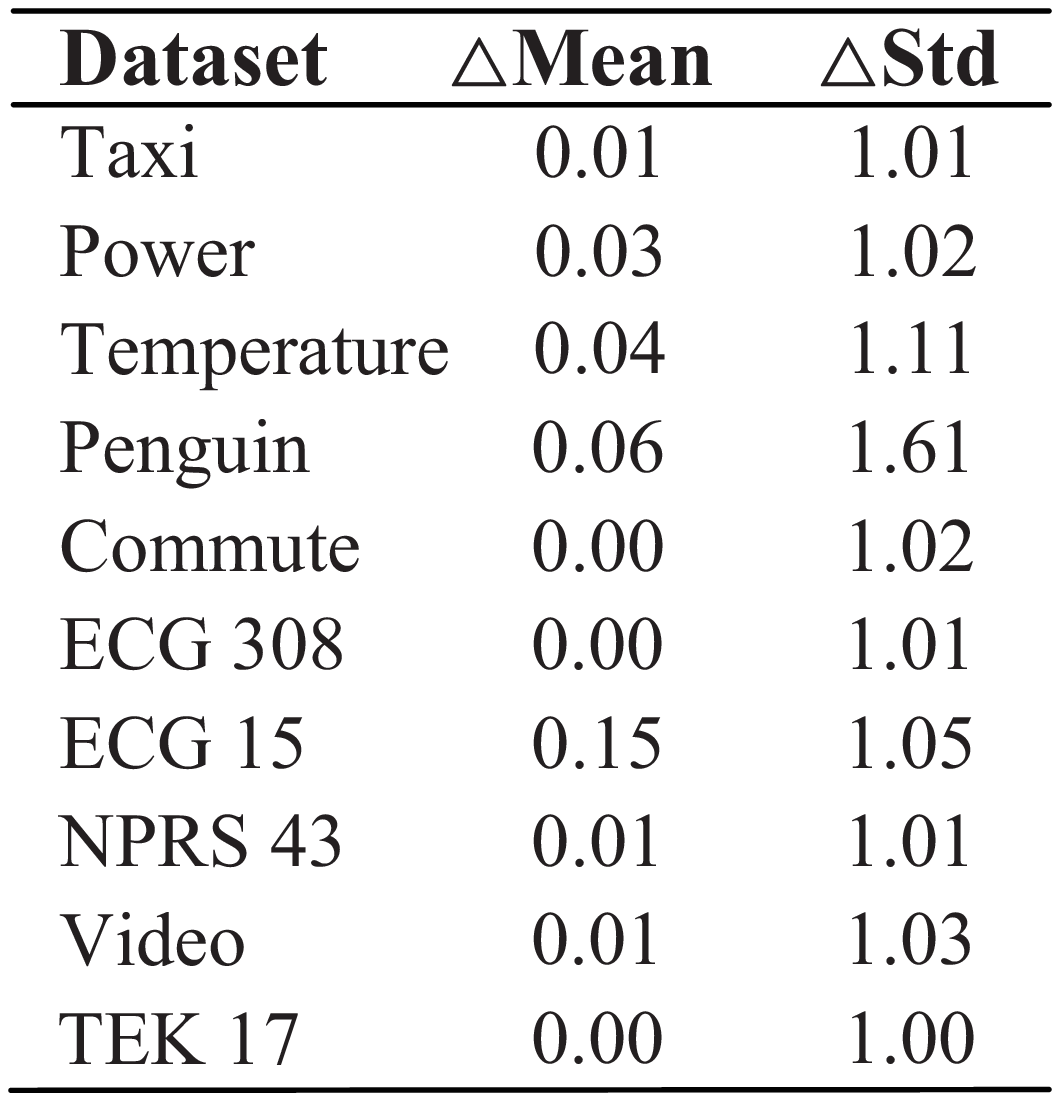}
		\caption{\label{fig:motif}Motif example}
	\end{minipage}%
\end{figure}

By observing Fig.~\ref{fig:cnsm2}(a), one can filter the undesired results easily by adding an additional constraint: the output subsequences should have similar mean value as $Q$. In fact, this new type of NSM query, \emph{NSM plus some constraints}, is useful in many applications. We list two of them as follows,
\begin{itemize}
\item{(Industry application)} In the wind power generation field, LIDAR system can provide preview information of wind disturbances~\cite{branlard2009wind}. Extreme Operating Gust (EOG) is a typical gust pattern which is a phenomenon of dramatic changes of wind speed in a short period. Fig.~\ref{fig:wind} shows a typical EOG pattern. This pattern is important because it may generate damage on the turbine. All EOG pattern occurrences have the similar shape, and their fluctuation degree falls within certain range, because the wind speed cannot be arbitrarily high.
    If we hope to find all EOG pattern occurrences in the historical data, we can use a typical EOG pattern as the query, plus the constraint on the range of the values.
\item{(IoT application)} When a container truck goes through a bridge, the strain meter planted in the bridge will demonstrate a specific fluctuation pattern. The value range in the pattern depends on the weight of the truck. If we have one occurrence of the pattern as a query, we can additionally set a mean value range as the constraint to search container trucks whose weight falls within a certain range.
\end{itemize}

Note that the above applications cannot be handled by RSM query, because the existing offset shifting and amplitude scaling forces us to set a very large distance threshold, which will cause many false positive results.

Furthermore, to verify the universality of this new query type, we investigate the motif pairs in some popular real-world time series benchmarks.
Motif mining~\cite{matrix_profile} is an important time series mining task, which finds a pair (or set) of subsequences with minimal normalized distance. For a motif subsequence pair, say $X$ and $Y$, we show the relative mean value difference ($\Delta$Mean$=\frac{|\mu^X-\mu^{Y}|}{\max-\min}$) and the ratio of standard deviation ($\Delta$Std$=|\frac{\sigma^X}{\sigma^Y}|$) in Fig.~\ref{fig:motif}
We can see that although these pairs are found without any constraint (like NSM query), both mean value and standard deviation of motif subsequences are very similar. So we can find these pairs by the cNSM query, a NSM query plus a small constraint.

In this paper, we formally define a new subsequence matching problem, called \emph{constrained normalized subsequence matching problem} (cNSM for short). Two constraints, one for mean value and the other for standard deviation, are added to the traditional NSM problem. One exemplar cNSM query looks like ``given a query $Q$ with mean value $\mu^Q$ and standard deviation $\sigma^Q$, return subsequences $S$ which satisfy: (1) $Dist(\hat{S},\hat{Q})\leq 1.5$; (2) $|\mu^Q-\mu^S|\leq 5$; (3) $0.5\leq \sigma^Q/\sigma^S\leq 2$''. With the constraint, the cNSM problem provides a knob to flexibly control the degree of offset shifting (represented by mean value) and amplitude scaling (represented by standard deviation). Moreover, the cNSM problem offers us the opportunity to build index for the normalized subsequence matching.

\textbf{Challenges.} Solving the cNSM problem faces the following challenges. First, how can we process the cNSM query efficiently? A straightforward approach is to first apply UCR Suite to find unconstrained results, and then use mean value and standard deviation constraints to prune the unqualified ones. However, it still needs to scan the full series. Can we build an index and process the query more efficiently?

Second, users often conduct the similar subsequence search in an exploratory and interactive fashion. Users may try different distance functions, like Euclidean distance or Dynamic Time Warping. Meanwhile, users may try RSM and cNSM query simultaneously. Can we build a single index to support all these query types?

\textbf{Contributions.}
Besides proposing the cNSM problem, we also have the following contributions.
\begin{itemize}
\item We present the filtering conditions for four query types, RSM-ED, RSM-DTW, cNSM-ED and cNSM-DTW, and prove the correctness. The conditions enable us to build index and meanwhile guarantee no false dismissals.
\item We propose a new index structure, KV-index, and the query processing approach, KV-match, to support all these query types. The biggest advantage is that we can process various types of queries efficiently with a single index. Moreover, KV-match only needs a few numbers of \emph{sequential scans} of the index, instead of many random accesses of tree nodes in the traditional R-tree index, which makes it much more efficient.
\item Third, to support the query of arbitrary lengths efficiently, we extend KV-match to KV-match$_{\text{DP}}$, which utilizes multiple indexes with different window lengths. We conduct extensive experiments. The results verify the efficiency and effectiveness of our approach.
\end{itemize}

The rest of the paper is organized as follows. We present the preliminary knowledge and problem statements in Section~\ref{sec:preliminary}. In Section~\ref{sec:motivation} we introduce the theoretical foundation and motivate the approach. Section~\ref{sec:index} and \ref{sec:matching} describe our index structure, index building algorithm and query processing algorithm. Section~\ref{sec:dynamic} extends our method to use multi-level indexes with different window lengths. Our implementation details are described in Section~\ref{sec:implementation}.
The experimental results are presented in Section~\ref{sec:experiment} and we discuss related works in Section~\ref{sec:related}. Finally, we conclude the paper and look into the future work in Section~\ref{sec:conclusion}.

%% file: tex/preliminary1.tex
\section{Preliminary Knowledge}

\label{sec:preliminary}

In this section, we introduce the definition of time series and other useful notations.

\subsection{Definitions and Problem Statement}

\begin{table}[tbp]
	\centering
	\label{tab:notation}
	\caption{Frequently used notations}
	\begin{tabular}{ll}
		\hline
		Notation         & Description                                            \\
		\hline
		$X$              & a time series $(x_1, x_2, \cdots, x_n)$                \\
		$X(i,l)$         & a length-$l$ subsequence of $X$ starting at offset $i$ \\
		$\hat{X}$        & the normalized series of time series $X$               \\
		$X_i$            & the $i$-th length-$w$ disjoint window of $X$                      \\
		$\mu^X_i $       & the mean value of the $i$-th disjoint window of $X$             \\
		$\sigma^X_i$     & the standard deviation of the $i$-th disjoint window of $X$     \\
		$\textit{WI}$       & a window interval containing continuous window positions \\
		$\textit{IS}_i$     & a set of window intervals satisfying the criterion for $Q_i$ \\
		$\textit{CS}_i, \textit{CS}$ & a set of candidates for $Q_i$ and for all $Q_j (1 \leq j \leq i)$\\
		$n_I, n_P$ & the number of window intervals and window positions \\ 
		\hline
	\end{tabular}
\end{table}

A \textit{time series} is a sequence of ordered values, denoted as $X=(x_1, x_2, \cdots, x_n)$, where $n=|X|$ is the \textit{length} of $X$. A length-$l$ \textit{subsequence} of $X$ is a shorter time series, denoted as $X(i,l) = (x_i,x_{i+1},\cdots,x_{i+l-1})$, where $1 \leq i \leq n-l+1$.

For any subsequence $S=(s_1,s_2,\cdots,s_m)$, $\mu^{S}$ and $\sigma^{S}$ are the \textit{mean value} and \textit{standard deviation} of $S$ respectively. Thus the \textit{normalized series} of $S$, denoted as $\hat{S}$, is
\small
\begin{equation}
\label{eq:normaliedseries}
\hat{S}=\left(\frac{s_1-\mu^{S}}{\sigma^{S}},\frac{s_{2}-\mu^{S}}{\sigma^{S}},\cdots,\frac{s_{m}-\mu^{S}}{\sigma^{S}}\right)
\nonumber
\end{equation}
\normalsize

Our work supports two common distance measures, \textit{Euclidean distance} and \textit{Dynamic Time Warping}.
Here we give the definition of them.

\textit{Euclidean Distance} (\textbf{ED}):
Given two length-$m$ sequences, $S$ and $S'$, their distance is
${\textit{ED}}(S,S')=\sqrt{\textstyle \sum_{i=1}^{m} (s_i - s'_i)^{2}}\nonumber$.

\textit{Dynamic Time Warping} (\textbf{DTW}):
Given two length-$m$ sequences, $S$ and $S'$, their distance is
\small
\begin{multline}
{\textit{DTW}}(\left<\right>,\left<\right>)=0 ;~~~~~ {\textit{DTW}}(S,\left<\right>)={\textit{DTW}}(\left<\right>,S')=\infty \\
{\textit{DTW}}(S,S')=\sqrt{(s_{1}-s'_{1})^{2} + \min
	\left\{
	\begin{aligned}
	& {\textit{DTW}}(suf(S),suf(S'))\\
	& {\textit{DTW}}(S,suf(S'))\\
	& {\textit{DTW}}(suf(S),S')
	\end{aligned}
	\right.
}\nonumber
\end{multline}
\normalsize
where $\left<\right>$ represents empty series and $suf(S)=(s_2,\cdots,s_m)$ is a suffix subsequence of $S$.

In DTW, the \textit{warping path} is defined as a matrix to represent the optimal alignment for two series. The matrix element $(i,j)$ represents that $s_i$ is aligned to $s'_j$. To reduce the computation complexity, we use the Sakoe-Chiba band~\cite{sakoe} to restrict the width of warping, denoted as $\rho$. Any pair $(i,j)$ should satisfy $|i-j|\leq \rho$. When $\rho = 0$, it degenerates into ED.

%

We aim to support subsequence matching for both the raw subsequence and the normalized subsequence simultaneously. The problem statements are given here.

\textit{Raw Subsequence Matching} (\textbf{RSM}):
Given a long time series $X$, a query sequence $Q$ ($|X|\geq|Q|$) and a distance threshold $\varepsilon ~(\varepsilon \geq 0)$, find all subsequences $S$ of length $|Q|$ from $X$, which satisfy
$D\big(S,Q\big) \leq \varepsilon$.
In this case, we call that $S$ and $Q$ are \textit{in $\varepsilon$-match}.

\textit{Normalized Subsequence Matching} (\textbf{NSM}):
Given a long time series $X$, a query sequence $Q$ and a distance threshold $\varepsilon ~(\varepsilon \geq 0)$, find all subsequences $S$ of length $|Q|$ from $X$, which satisfy
$D\big(\hat{S},\hat{Q}\big) \leq \varepsilon$,
where $\hat S$ and $\hat Q$ are the normalized series of $S$ and $Q$ respectively.

The cNSM problem adds two constraints to the NSM problem. Thresholds $\alpha~(\alpha \geq 1)$ and $\beta~(\beta \geq 0)$ are introduced to constrain the degree of amplitude scaling and offset shifting.

\textit{Constrained Normalized Subsequence Matching} (\textbf{cNSM}):
Given a long time series $X$, a query sequence $Q$, a distance threshold $\varepsilon$, and the constraint thresholds $\alpha$ and $\beta$, find all subsequences $S$ of length $|Q|$ from $X$, which satisfy
\begin{equation}
D\big(\hat{S},\hat{Q}\big) \leq \varepsilon~,~
\frac{1}{\alpha} \leq \frac{\sigma^S}{\sigma^Q} \leq \alpha~,~
-\beta \leq \mu^S-\mu^Q \leq \beta
\nonumber
\end{equation}
The larger $\alpha$ and $\beta$, the looser the constraint. In this case, we call that $S$ and $Q$ are \textit{in $(\varepsilon,\alpha,\beta)$-match}.


The distance $D(\cdot,\cdot)$ is either ED or DTW. In this paper, we build an index to support four types of queries, RSM-ED, RSM-DTW, cNSM-ED and cNSM-DTW simultaneously.

%% file: tex/theoretical.extended.tex
\needspace{10mm}
\section{Theoretical Foundation and \\Approach Motivation}
\label{sec:motivation}
In this section, we establish the theoretical foundation of our approach. We propose a condition to filter the unqualified subsequences. For all four types of queries, the conditions share the same format, which enables us to support all query types with a single index.

Specifically, for the query $Q$ and the subsequence $S$ of length-$m$, we segment them into aligned disjoint windows of the same length $w$. The $i$-th window of $Q$ (or $S$) is denoted as $Q_i$ (or $S_i$), ($1\leq i \leq p=\left\lfloor\frac{m}{w}\right\rfloor$), that is, $Q_i=(q_{(i-1)* w+1},\cdots,q_{i* w})$.

For each window, we hope to find one or more features, based on which we can construct the filtering condition. In this work, we choose to utilize one single feature, the mean value of the window. The advantages are two-folds. First, with a single feature, we can build a one-dimensional index, which improves the efficiency of index retrieval greatly. Second, the mean value allows us to design the condition for both RSM and cNSM query.

We denote mean values of $Q_i$ and $S_i$ as $\mu_i^Q$ and $\mu_i^S$. The condition consists of $p$ number of ranges. The $i^{th}$ one is denoted as $[\textit{LR}_i,\textit{UR}_i]$ ($1\leq i \leq p$). If $S$ is a qualified subsequence, for any $i$, $\mu_i^S$ must fall within $[\textit{LR}_i,\textit{UR}_i]$. If any $\mu_i^S$ is outside the range, we can filter $S$ safely.


\subsection{RSM-ED Query Processing}
\label{section:RSM}
In this section, we first present the condition for the simplest case, RSM-ED query, and then illustrate our approach.


\begin{lemma}
	\label{lemma:RSM-ED}
	If $S$ and $Q$ are in $\varepsilon$-match under \textit{ED} measure, that is, $\textit{ED}(S,Q)\leq \varepsilon$, then $\mu_i^S~(1\leq i\leq p)$ must satisfy
	\begin{equation}
	\mu^{S}_{i} \in \left[\mu^{Q}_{i} - \frac{\varepsilon}{\sqrt{w}}, \mu^{Q}_{i} + \frac{\varepsilon}{\sqrt{w}}\right]
    \label{eq:RSM-ED}
	\end{equation}
	\proof
	Based on the ED definition, we have
	\begin{equation}
	\textit{ED}^2\left(S, Q\right) = \sum_{k=1}^{n}\left({s_k} - {q_k}\right)^{2} \geq
	\sum_{j=(i-1)* w +1}^{i* w}\left({s_j} - {q_j}\right)^{2} \nonumber
	\end{equation}
	where $1 \leq i \leq p$. According to the corollary in $\cite{vldb00}$,
	\begin{equation}
	\sum_{j=(i-1)* w +1}^{i* w}\left({s_j} - {q_j}\right)^{2} \geq
	{w} * \left(\mu^{S}_{i} - \mu^{Q}_{i}\right)^{2} \nonumber
	\end{equation}
	If $D\left(S, Q\right) \leq \varepsilon$, after inequality transformation, it should hold that $\left(\mu^{S}_{i} - \mu^{Q}_{i}\right)^{2} \leq \frac{\varepsilon^2}{w}$, so we get Eq.~(\ref{eq:RSM-ED}).
	\qed
\end{lemma}

\begin{figure}[t]
	\centering
	\includegraphics[width=\linewidth]{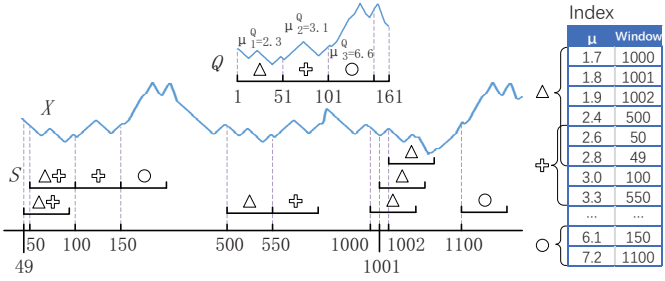}
	\caption{Illustrative example}
	\label{fig:example}
\end{figure}

Now we illustrate our approach with the example in Fig.~\ref{fig:example}. $X$ is a long time series, and $Q$ is the query sequence of length $161$. The goal is to find all length-$161$ subsequences $S$ from $X$, which satisfy $\textit{ED}(S,Q)\leq \varepsilon$. The parameter of the window length $w$ is set to 50. We split $Q$ into three disjoint windows of length $50$, $Q_1$, $Q_2$, $Q_3$\footnote{We can ignore the remain part $Q(151, 11)$ without sacrificing the correctness since Lemma~\ref{lemma:RSM-ED} is a \textit{necessary condition} for RSM.}. According to Lemma~\ref{lemma:RSM-ED}, for any qualified subsequence $S$, the mean value of the $i^{th}$ disjoint window $S_i$ must fall within the range $\left[\mu^{Q}_{i} - \frac{\varepsilon}{\sqrt{50}}, \mu^{Q}_{i} + \frac{\varepsilon}{\sqrt{50}}\right]$ ($i=1,2,3$). To facilitate finding the windows satisfying this condition, we build the index as follows. We compute the mean values of all sliding windows $X(j,w)$, denoted as $\mu(X(j,w))$, and build a sorted list of $\left<\mu(X(j,w)),j\right>$ entries. With this structure, we find the candidates in two steps. First, for each window $Q_i$, we obtain all sliding windows whose mean values fall within $\left[\mu^{Q}_{i} - \frac{\varepsilon}{\sqrt{50}}, \mu^{Q}_{i} + \frac{\varepsilon}{\sqrt{50}}\right]$ by a single \emph{sequential scan} operation. We denote the found windows for $Q_i$ as $\textit{CS}_i$. Then, we generate the final candidates by \emph{intersecting} windows in $\textit{CS}_1$, $\textit{CS}_2$ and $\textit{CS}_3$.

In Fig.~\ref{fig:example}, sliding windows in $\textit{CS}_1$, $\textit{CS}_2$ and $\textit{CS}_3$ are marked with ``triangle'', ``cross'' and ``circle'' respectively. The only candidate is $X(50,161)$, because  $X(50,50)\in \textit{CS}_1$, $X(100,50)\in \textit{CS}_2$ and $X(150,50)\in \textit{CS}_3$. 

\subsection{Range for cNSM-ED Query}
\label{section:cNSM}
%

We solve the cNSM problem based on KV-index either. For the given query $Q$, we determine whether a subsequence $S$ is $(\varepsilon, \alpha, \beta)$-match with $Q$ by checking the raw subsequence $S$ directly. Specifically, we achieve this goal by designing the range $[LR_i, UR_i]$ for each query window $Q_i$. For any subsequence $S$, if any $\mu^S_i$ falls outside this range, $S$ cannot be $(\varepsilon, \alpha, \beta)$-match with $S$ and we can filter $S$ safely. We illustrate it with an example. Let $Q=(1,1,-1,-1)$, $w=2$, $(\alpha,\beta)=(2,1)$ and $\varepsilon=0$~\footnote{To make the example simple enough, we set $\varepsilon$ as 0.}. By simple calculation, we obtain $\mu^Q_1=1$ and $\sigma^Q=1.1547$. For any length-4 subsequence $S$, if only $\mu^S_1=4$, we can infer that $S$ cannot be matched with $Q$ without checking the whether $\hat{S}$ satisfies the cNSM condition, as follows. To make $ED(\hat{Q},\hat{S})=0$, $\mu^S_2$ must be -4. If it is the case, $\sigma^S$ is 4.6188 at least. However, $\frac{\sigma^S}{\sigma^Q}>2$, which violates the cNSM condition.

%


Now we formally give the range for cNSM-ED query. Let $\mu^{S}$ and $\mu^{Q}$ be the global mean values of $S$ and $Q$, $\sigma^{S}$ and $\sigma^{Q}$ be the standard deviations, $\hat{S}$ and $\hat{Q}$ be the normalized $S$ and $Q$ respectively.

\begin{figure}[t]
	\begin{minipage}[t]{0.4\linewidth}
		\centering
		\includegraphics[width=0.93\linewidth]{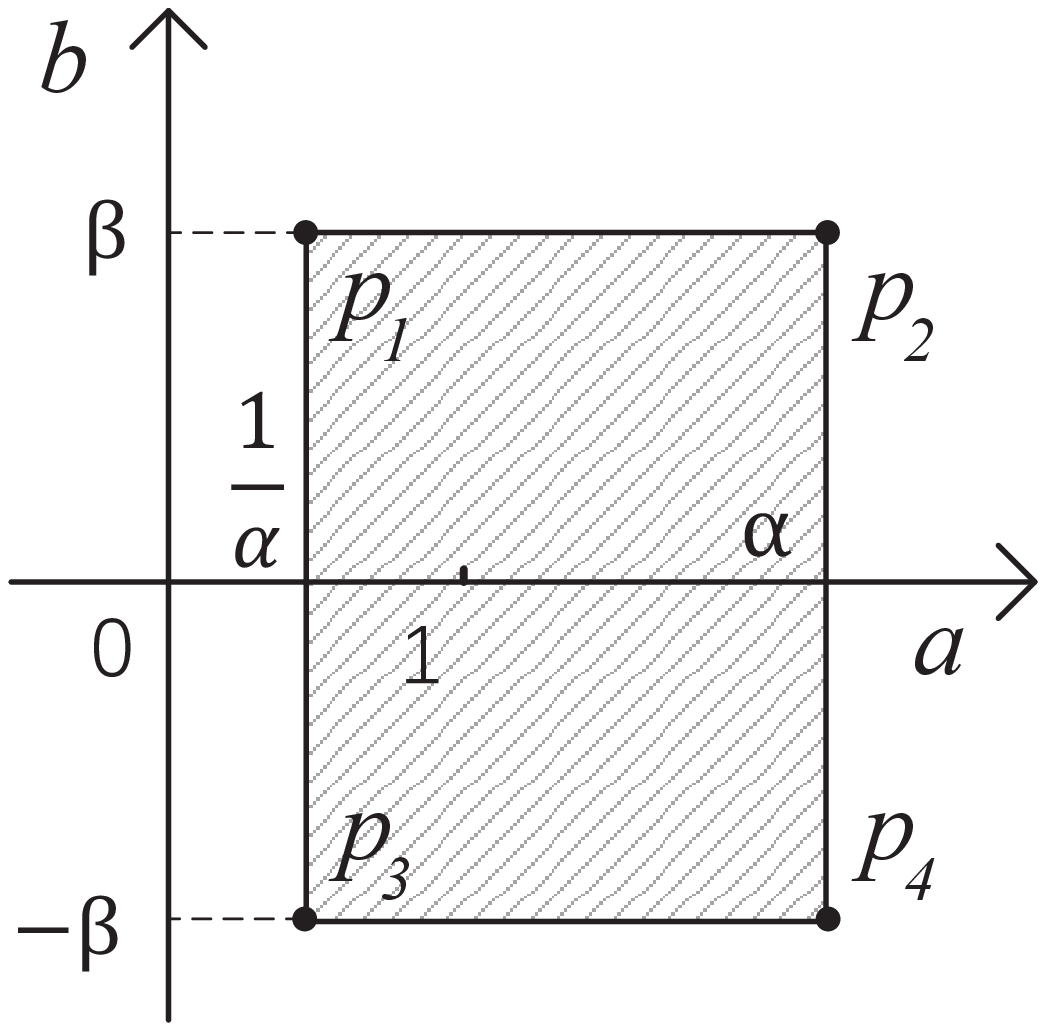}
		\caption{\label{fig:range}Legal Range of $(a,b)$}
	\end{minipage}%
	\hfill%
	\begin{minipage}[t]{0.58\linewidth}
		\centering
		\includegraphics[width=0.9\linewidth]{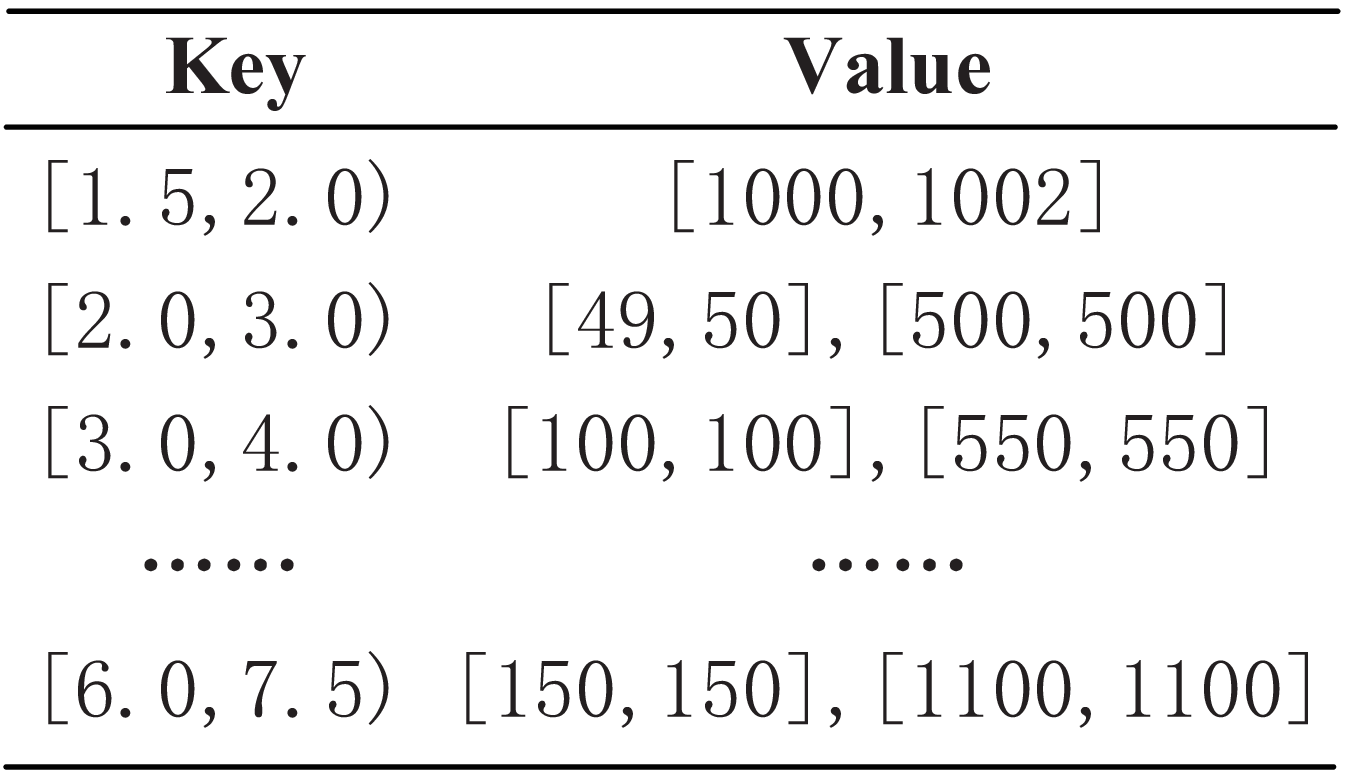}
		\caption{\label{fig:index}Index Structure}
	\end{minipage}%
\end{figure}

\begin{lemma}
\label{lemma:cNSM-ED}
	If $S$ and $Q$ are in $(\varepsilon, \alpha, \beta)$-match under \textit{ED} measure, that is, $\textit{ED}(\hat{S},\hat{Q})\leq \varepsilon$, then $\mu^S_i~(1\leq i\leq p)$ satisfies
\begin{equation}
\mu_i^S \in \left[v_{\min}+\mu^Q-\beta, v_{\max}+\mu^Q+\beta\right]
\label{eq:lemma3}
\end{equation}
where\\
$v_{\min}= \tiny{~} \min \left(\alpha\cdot(\mu_i^Q-\mu^Q- \frac{\varepsilon  \sigma^Q}{\sqrt{w}}), \frac{1}{\alpha}\cdot(\mu_i^Q-\mu^Q- \frac{\varepsilon  \sigma^Q}{\sqrt{w}}) \right)$,\\
$v_{\max}= \max \left(\alpha\cdot(\mu_i^Q-\mu^Q+ \frac{\varepsilon  \sigma^Q}{\sqrt{w}}), \frac{1}{\alpha}\cdot(\mu_i^Q-\mu^Q+ \frac{\varepsilon  \sigma^Q}{\sqrt{w}}) \right)$.

	\proof
	Based on the normalized ED definition, we have
	\small
	\begin{equation}
	\textit{ED}\big(\hat{S}, \hat{Q}\big) = \sqrt{\sum_{j=1}^{m}\left({\frac{s_j-\mu^S}{\sigma^S}} - {\frac{q_j-\mu^Q}{\sigma^Q}}\right)^{2}} \nonumber
	\end{equation}
	\normalsize
	
	Let $a=\frac{\sigma^{S}}{\sigma^{Q}}$ and $b=\mu^S-\mu^Q$, where $a \in [\frac{1}{\alpha}, \alpha]$ and $b \in [-\beta,\beta]$. If $\textit{ED}\big(\hat{S}, \hat{Q}\big) \leq \varepsilon$, it holds that
	\small
	\begin{equation}
	\sum_{j=1}^{m}\left({\frac{s_j-\mu^Q-b}{a \sigma^Q}} - {\frac{q_j-\mu^Q}{\sigma^Q}}\right)^{2} \leq \varepsilon^2  \nonumber
	\end{equation}
	\normalsize
	
	According to the corollary in $\cite{vldb00}$, similar to Lemma~\ref{lemma:RSM-ED}, for the $i$-th window $S_i$ and $Q_i$, we have
	\small
	\begin{equation}
	\left({\frac{\mu_i^S-\mu^Q-b}{a \sigma^Q}} - {\frac{\mu_i^Q-\mu^Q}{\sigma^Q}}\right)^{2} \leq \frac{\varepsilon^2}{w}  \nonumber
	\end{equation}
	\normalsize
	
    By simple transformation, for any specific pair of $(a,b)$, we can get a range of $\mu_i^S$ as follows,
	\small
	\begin{equation}
	\mu_i^S \in  \bigg[\left(\mu_i^Q-\mu^Q-\frac{\varepsilon  \sigma^Q}{\sqrt{w}}\right) a + b + \mu^Q,  \left(\mu_i^Q-\mu^Q+\frac{\varepsilon  \sigma^Q}{\sqrt{w}}\right) a + b + \mu^Q\bigg] \nonumber
	\end{equation}
	\normalsize
	
For ease of description, we assign $\mu_i^Q-\mu^Q-\frac{\varepsilon \sigma^Q}{\sqrt{w}}=A$ and $\mu_i^Q-\mu^Q+\frac{\varepsilon \sigma^Q}{\sqrt{w}}=B$.
	
	The final range $[\textit{LR}_i, \textit{UR}_i]$ should be
	\begin{flalign*}
	\Bigg[\min\limits_{
		\substack{a \in [\frac{1}{\alpha},\alpha]\\
			b \in [-\beta,\beta]}}
	\left\{A  a + b + \mu^Q\right\},
	\max\limits_{
		\substack{a \in [\frac{1}{\alpha},\alpha]\\
			b \in [-\beta,\beta]}}
	\left\{B  a + b + \mu^Q\right\}
	\Bigg]
	\end{flalign*}
	
	As illustrated in Fig.~\ref{fig:range}, the rectangle represents the whole legal range of $a$ and $b$. Let $f(a,b)=A a+b+\mu^Q$ and $g(a,b)=B a+b+\mu^Q$. Apparently, both $f(a,b)$ and $g(a,b)$ increase monotonically for $b \in [-\beta,\beta]$. As for $a$, we have two cases,
	\begin{itemize}
		\item If $A \ge 0$, $f(a,b)$ increases monotonically for $a \in [\frac{1}{\alpha},{\alpha}]$. $f(a,b)$ is minimal when $a={\frac{1}{\alpha}}$ and $b={-\beta}$, which is represented by the point $p_3$ in Fig.~\ref{fig:range};
		\item If $A<0$, $f(a,b)$ decreases monotonically for $a \in [\frac{1}{\alpha},{\alpha}]$. $f(a,b)$ is minimal when $a={\alpha}$ and $b={-\beta}$, which is represented by the point $p_4$ in Fig.~\ref{fig:range}.
	\end{itemize}
	\begin{flalign*}
	\text{So} && \textit{LR}_i=\min\limits_{a \in [\frac{1}{\alpha},\alpha],
		b \in [-\beta,\beta]} f(a,b) &= \min\limits_{a \in \left\{\frac{1}{\alpha},\alpha \right\}} f(a,-\beta) &
	\end{flalign*}

    Note that formula $a \in \left\{\frac{1}{\alpha},\alpha\right\}$ means $a$ is either $\frac{1}{\alpha}$ or $\alpha$.
	
	Similarly, we can infer the maximal value of $g(a,b)$ as following two cases,
	\begin{itemize}
		\item If $B \ge 0$, $g(a,b)$ is maximal when $a={\alpha}$ and $b={\beta}$, which is represented by the point $p_2$ in Fig.~\ref{fig:range}.
		\item If $B<0$, $g(a,b)$ is maximal when $a={\frac{1}{\alpha}}$ and $b={\beta}$, which is represented by the point $p_1$ in Fig.~\ref{fig:range}.
	\end{itemize}
	\begin{flalign*}
	\text{So} && \textit{UR}_i=\max\limits_{a \in [\frac{1}{\alpha},\alpha],
		b \in [-\beta,\beta]} g(a,b) &= \max\limits_{a \in \left\{\frac{1}{\alpha},\alpha \right\}} g(a,\beta) &
	\end{flalign*}
	\qed
\end{lemma}

\vspace{-7mm}
\subsection{Range for RSM-DTW and cNSM-DTW Query}
Before introducing the ranges, we first review the query envelop and the lower bound of DTW distance, LB\_PAA~\cite{lb_paa}. To deal with DTW$_{\rho}$ measure, given length-$m$ query $Q$, the query envelop consists of two length-$m$ series, $L$ and $U$, as the lower and upper envelop respectively. The $i$-th elements of $L$ and $U$, denoted as $l_i$ and $u_i$, are defined as
\begin{center}
	$\displaystyle l_i=\min_{-\rho\leq r\leq \rho} q_{i+r}$ , $\displaystyle u_i=\max_{-\rho\leq r\leq \rho} q_{i+r}$.
\end{center}

LB\_PAA is defined based on the query envelop. $L$ and $U$ are split into $p$ number of length-$w$ disjoint windows, $(L_1,L_2,\cdots,L_p)$ and $(U_1,U_2,\cdots,U_p)$, in which $L_i=(l_{(i-1)\cdot w+1},\cdots,l_{i\cdot w})$ and $U_i=(u_{(i-1)\cdot w+1},\cdots,u_{i\cdot w})$ ($1\leq i \leq p=\lfloor \frac{m}{w}\rfloor$). The mean values of $L_i$ and $U_i$ are denoted as $\mu_i^L$ and $\mu_i^U$ respectively. For any length-$m$ subsequence $S$, the LB\_PAA is as follows,

\vspace{-4mm}
\small
\begin{equation}
\textit{LB\_PAA}(S,Q)=
\sqrt{\sum_{i=1}^{p}w\cdot
	\begin{lcases}
	&(\mu_i^S-\mu_i^U)^2 && \text{if~} \mu_i^S > \mu_i^U \\
	&(\mu_i^S-\mu_i^L)^2 && \text{if~} \mu_i^S < \mu_i^L \\
	&0 && \text{Otherwise} \\
	\end{lcases}
}
\label{eq:lb_paa}
\end{equation}
\normalsize
which satisfies $\textit{LB\_PAA}(S,Q)\leq \textit{DTW}_{\rho}(S,Q)$~\cite{lb_paa}.



Now we give the ranges for RSM and cNSM under the DTW$_\rho$ measure in turn.
\begin{lemma}
	\label{lemma:RSM-DTW}
	If $S$ and $Q$ are in $\varepsilon$-match under \textit{DTW}$_\rho$ measure, that is, $\textit{DTW}_\rho (S,Q)\leq \varepsilon$, then $\mu_i^S~(1\leq i\leq p)$ satisfies
	\begin{equation}
	\mu^{S}_{i} \in \left[\mu^{L}_{i} - \frac{\varepsilon}{\sqrt{w}}, \mu^{U}_{i} + \frac{\varepsilon}{\sqrt{w}}\right]
    \label{eq:RSM-DTW}
	\end{equation}
	\proof See Appendix~\ref{app:1}.
\end{lemma}


\begin{lemma}
	If $S$ and $Q$ are in $(\varepsilon, \alpha, \beta)$-match under \textit{DTW}$_\rho$ measure, that is, $\textit{DTW}_\rho(\hat{S},\hat{Q})\leq \varepsilon$, then $\mu^S_i~(1\leq i\leq p)$ satisfies
	\label{lemma:cNSM-DTW}
	\begin{equation}
	\mu_i^S \in \left[v_{\min}+\mu^Q-\beta, v_{\max}+\mu^Q+\beta\right]
    \label{eq:lemma4}
	\end{equation}
	where\\
	$v_{\min}=\tiny{~} \min \left(\alpha\cdot(\mu_i^L-\mu^Q- \frac{\varepsilon  \sigma^Q}{\sqrt{w}}), \frac{1}{\alpha}\cdot(\mu_i^L-\mu^Q- \frac{\varepsilon  \sigma^Q}{\sqrt{w}}) \right)$,\\
	$v_{\max}= \max \left(\alpha\cdot(\mu_i^U-\mu^Q+ \frac{\varepsilon  \sigma^Q}{\sqrt{w}}), \frac{1}{\alpha}\cdot(\mu_i^U-\mu^Q+ \frac{\varepsilon  \sigma^Q}{\sqrt{w}}) \right)$.
	\proof See Appendix~\ref{app:2}.
\end{lemma}

\textbf{Analysis}. We provide the ranges of mean value for all four query types, which means that we can support all queries with a single index. When processing different query types, the only difference is to use different ranges of $\mu_{i}^S$. This property is beneficial for exploratory search tasks.


%% file: tex/index.extended.tex
\section{KV-\MakeLowercase{index}}
\label{sec:index}
In this section, we present our index structure KV-index, and the index building algorithm.

\subsection{Index Structure}
\label{sec:index_structure}

The index structure in Fig.~\ref{fig:example} has approximately equal number of entries of $|X|$, which causes a huge space cost. To avoid that, we propose a more compact index structure which utilizes the data locality property, that is, the values of adjacent time points may be close. In consequence, the mean values of adjacent sliding windows will be similar too.

Logically, KV-index consists of ordered rows of key-value pairs. The key of the $i$-th row, denoted as $K_i$, is a \emph{range} of mean values of sliding windows, that is,
$K_i=[low_i, up_i)$,
where $low_i$ and $up_i$ are the left and right endpoint of the mean value range of $K_i$ respectively. It is a left-closed-right-open range, and the ranges of adjacent rows are disjoint.

The corresponding value, denoted as $V_i$, is the set of sliding windows whose mean values fall within $K_i$. To facilitate the expression, we represent each window by its position, that is, we represent sliding window $X(j, w)$ with $j$. To further save the space cost and also facilitate subsequence matching algorithm, we organize the window positions in $V_i$ as follows. The positions in $V_i$ are sorted in ascending order, and consecutive ones are merged into a \emph{window interval}, denoted as $\textit{WI}$. So $V_i$ consists of one or more sorted and non-overlapped window intervals.

\begin{definition}[Window Interval]
	We combine the $l^{th}$ to $r^{th}$ length-$w$ sliding windows of $X$ as a window interval $\textit{WI} = [l, r]$, which contains a set of sliding windows $\{X(l,w), X(l+1,w), \cdots, X(r,w)\}$, where $1 \leq l \leq r \leq |X|-w+1$.
\end{definition}

In the following descriptions, we use $j \in \textit{WI}$ to denote the window position $j$ belonging to the window interval $\textit{WI}=[l,r]$, that is, $j \in [l, r]$. Moreover, we use $\textit{WI}.l$, $\textit{WI}.r$ and $|\textit{WI}|=r-l+1$ to denote the left boundary, the right boundary and the size of interval $\textit{WI}$ respectively. The overall number of window intervals in $V_i$ is denoted as $n_I(V_i)$, and the number of window positions in $V_i$ as $n_P(V_i)$. Formally, we have
\begin{align}
\label{eq:niv}
n_I(V_i) ={}& \big|\{\textit{WI}~|\textit{WI}\in V_i\}\big| \\
\label{eq:nov}
n_P(V_i) ={}& \sum_{\textit{WI}\in V_i} \big|\textit{WI}\big|
\end{align}

Fig.~\ref{fig:index} shows KV-index for Fig.~\ref{fig:example}. The first row indicates that there exists three sliding windows, $X(1000, 50)$, $X(1001, 50)$ and $X(1002, 50)$, whose mean values fall within the range $[1.5, 2.0)$. In the second row, three windows are organized into two intervals $[49, 50]$ and $[500,500]$. Thus $n_I(V_2)=2$ and $n_P(V_2)=3$. Note that, $[500,500]$ is a special interval which only contains one single window position.


To facilitate the query processing, KV-index also contains a meta table, in which each entry is a quadruple as
$ \left<K_i, pos_i, n_I(V_i), n_P(V_i)\right>$,
where $pos_i$ is the offset of $i$-th row in the index file. Due to its small size, we can load the meta table to memory before processing the query. With the meta table, we can quickly determine the offset and the length of a scan operation by the simple binary search.

Physically, KV-index can be implemented as a local file, an HDFS file or an HBase table, because of its simple format. In this work, we implement two versions, a local file version and an HBase table version (details are in Section~\ref{sec:experiment}). In general, if a file system or a database supports the ``scan'' operation with start-key and end-key parameters, it can support KV-index. 
We provide details about the index implementation in Section~\ref{sec:implementation}.

\subsection{Index Building Algorithm}
\label{sec:index_building_algorithm}
We build the index with two steps. First, we build an index in which all rows use the equal-width range of the mean values. Second, because data distribution is not balanced among rows, we merge adjacent rows to optimize the index. We first introduce a basic in-memory algorithm, which works for moderate data size. Then we discuss how to extend it to very large data scale.

In the first step, we pre-define a parameter $d$, which represents the range width of the mean values. The range of each row will be $[k\cdot d, (k+1) \cdot  d)$, where $k \in \mathbb{Z}$. We read series $X$ sequentially. A circular array is used to maintain the length-$w$ sliding window $X(i,w)$, and its mean value $\mu^X_{i}$ are computed on the fly. Assume the mean value of $S_{i-1}$, $\mu^X_{i-1}$, is in range $K_j$, and the mean value of the current window $S_i$, $\mu^X_{i}$, is also in $K_j$, we modify the current $\textit{WI}$ by changing its right boundary from $i-1$ to $i$. Otherwise, a new interval, $\textit{WI}=[i,i]$, will be added into certain row according to $\mu_i^X$.

The equal-width range can cause the zigzag style of adjacent rows. For example, the $V_i=\{[5,5], [7,7]\}$ and $V_{i+1}=\{[6,6],[8,8]\}$. Apparently, a better way is to merge these two rows so that the corresponding value becomes $V_i=[5,8]$.

In the second step, we merge adjacent rows with a greedy algorithm. We check the rows beginning from $\langle K_1, V_1\rangle$ and $\langle K_2, V_2\rangle$. Let the current rows be $\langle K_i, V_i\rangle$ and $\langle K_{i+1}, V_{i+1}\rangle$. The merging condition is whether $\frac{n_I(V_{i} \cup V_{i+1})}{n_I(V_{i}) + n_I(V_{i+1})}$ is smaller than $\gamma$, a pre-defined parameter. The rationale is that we merge the rows in which a large number of intervals are neighboring. If rows $\langle K_i, V_i\rangle$ and $\langle K_{i+1}, V_{i+1}\rangle$ are merged, the new key is $[low_i, up_{i+1})$, and the new value is $V_{i} \cup V_{i+1}$. Moreover, all neighboring window intervals from $V_i$ and $V_{i+1}$ are merged to one interval.


The merge operation is actually a union operation between two ordered interval sequences, which can be implemented efficiently similar to the merge-sort algorithm. Since each window interval will be examined exactly once, its time complexity is $O(n_I(V_{i})+n_I(V_{i+1}))$.

If the size of index exceeds memory capacity, we build the index as follows. In the first step, we divide time series into segments, and build the fixed-width range index for each segment in turn. After all segments are processed, we merge the rows of different segments. The second step visits index rows sequentially, which can be also divided into sub-tasks. Since each step can be divided into sub-tasks, the whole index building algorithm can be easily adapted to distributed environment, like MapReduce.

\textbf{Complexity analysis.} The process of building KV-index consists of two steps, generating rows with the fixed width, and merging them into varied-width ones. The first step scans all data in stream fashion, computes the mean value, and inserts $\left<\mu, offset\right>$ entry into hash table. Note that the mean value of $X(i,l)$ can be computed based on that of $X(i-1,l)$, whose cost is $O(1)$. So the cost of the first step is $O(n)$. In the second step, we detect adjacent rows and merge them if necessary. Since the intervals are ordered within each row, the merge operation is similar to the merge sort, whose cost is $n_I(V_i)+n_I(V_{i+1})$. Therefore, the whole cost is $\sum_{i=1}^{D-1} n_I(V_i)+n_I(V_{i+1})$ ($D$ is the number of rows in first step). Because $n_I(V_i)\leq n_P(V_i)$ and $\sum_{i=1}^{D} n_P(V_i) =n-w+1$, we can infer that its cost is $O(2n)$. In summary, the complexity of building index is $O(n)$.

All previous index-based approaches, like FRM and General Match, are based on R-tree, whose building cost is $O(n \cdot log_2(n))$~\cite{r-tree}. Moreover, they use DFT to transform each $w$-size window of $X$, whose cost is $w \cdot log_2(w)$. So the total transformation cost is $O(n \cdot w \cdot log_2(w))$. Therefore, building KV-index is more efficient.

%% file: tex/matching.tex
\section{KV-\MakeLowercase{match}}
\label{sec:matching}
In this section, we present the matching algorithm KV-match, whose pseudo-code is shown in Algorithm~\ref{alg:query}.

\subsection{Overview}
Initially, given query $Q$, we segment it into disjoint windows $Q_i$ of length $w$ ($1\leq i\leq p =\left\lfloor\frac{|Q|}{w}\right\rfloor$), and compute mean values $\mu_{i}^Q$ (Line \ref{alg1:initial}). We assume that $|Q|$ is an integral multiple of $w$. If not, we keep the longest prefix which is a multiple of $w$. According to the analysis in Section~\ref{sec:motivation}, the rest part can be ignored safely.

The main matching process consists of two phases:
\begin{enumerate}[{Phase} 1:]
	\item Index-probing (Line \ref{alg1:phase1:begin}-\ref{alg1:phase1:end}): For each window $Q_i$, we fetch a list of consecutive rows in KV-index according to the lemmas in Section~\ref{sec:motivation}. Based on these rows, we generate a set of subsequence candidates, denoted as $\textit{CS}$.
	\item Post-processing (Line \ref{alg1:phase2:begin}-\ref{alg1:phase2:end}): All subsequences in $\textit{CS}$ will be verified by fetching the data and computing the actual distance.
\end{enumerate}

Note that all four types of queries have the same matching process, the only difference is that in the index-probing phase, for each window, different types have the various row ranges, as introduced in Section~\ref{sec:motivation}.

\begin{algorithm}
	\small
	\caption{MatchSubsequence($X, w, Q, \varepsilon$)}
	\begin{algorithmic}[1]
		\State $p \gets \left\lfloor\frac{|Q|}{w}\right\rfloor, \mu_{i}^Q \gets \text{avg}(Q_i)~(1 \leq i \leq p)$ \label{alg1:initial}
		\For {$i \gets 1, p$} \label{alg1:phase1:begin}
			\State $\textit{RList}_i \gets \left\{\left<K_{s_i}, V_{s_i}\right>, \cdots, \left<K_{e_i}, V_{e_i}\right> \right\}$
			\label{alg1:phase1:sigma:begin}
			\State $\textit{IS}_i \gets \varnothing$
			\ForAll {$\left<K_j, V_j\right> \in \textit{RList}_i$}
				\State $\textit{IS}_i \gets \textit{IS}_i \cup \{\textit{WI}~|\textit{WI} \in V_j\}$
			\EndFor \label{alg1:phase1:sigma:end}
			\State $\Call{Sort}{\textit{IS}_i}$ \label{alg1:phase1:sort}
			\State $\textit{CS}_i \gets \varnothing, shift_i \gets (i-1) * w$ \label{alg1:phase1:shifting:sigma}
			\ForAll {$\textit{WI} \in \textit{IS}_i$} \label{alg1:phase1:shifting:begin}
				\State $\textit{CS}_i$.add($[\textit{WI}.l - shift_i, \textit{WI}.r - shift_i]$)
			\EndFor \label{alg1:phase1:shifting:end}
			\If {$i = 1$} $\textit{CS} = \textit{CS}_i$
			\Else {} $\textit{CS} \gets \Call{Intersect}{\textit{CS}, \textit{CS}_i}$ 
			\EndIf \label{alg1:phase1:intersection}
		\EndFor \label{alg1:phase1:end}
		\State $answers \gets \varnothing$ \label{alg1:phase2:begin}
		\ForAll {$\textit{WI} \in \textit{CS}$}
			\State $S \gets X(\textit{WI}.l, \textit{WI}.r-\textit{WI}.l+|Q|)$ \label{alg1:phase2:fetch} \Comment Scan from data
			\For {$j \gets 1, |S|-|Q|+1$}
				\If {$D(Q, S(j, |Q|)) \leq \varepsilon$} \Comment Extra test for cNSM \label{alg1:phase2:calculate}
					\State $answers$.add($S(j, |Q|)$)
				\EndIf
			\EndFor
		\EndFor \label{alg1:phase2:end}
		\State \Return $answers$
	\end{algorithmic}
	\label{alg:query}
\end{algorithm}

\subsection{Window Interval Generation}
For each window $Q_i$, we calculate the range of $\mu_i^S$, $[
\textit{LR}_i, \textit{UR}_i]$, firstly according to the query type. Then we visit KV-index with a single scan operation, which will obtain a list of consecutive rows, denoted as $\textit{RList}_i=\{\left<K_{s_i}, V_{s_i}\right>, \left<K_{s_i+1},\\ V_{s_i+1}\right>, \cdots, \left<K_{e_i}, V_{e_i}\right> \}$, which satisfies $\textit{LR}_i \in [low_{s_i},up_{s_i})$ and $\textit{UR}_i \in [low_{e_i},up_{e_i})$. Note that the $s_i$-th row (or the $e_i$-th row) may contain mean values out of the range. However, it only brings negative candidates, without missing any positive one.

We denote all window intervals in $\textit{RList}_i$ as $\textit{IS}_i=\{\textit{WI}~| \textit{WI}\in V_k, k \in [s_i, e_i]\}$. We use $\textit{WI} \in \textit{IS}_i$ to indicate that window interval $\textit{WI}$ belongs to $\textit{IS}_i$. Also, for any window position $j$ in $\textit{WI}$ ($\textit{WI}\in \textit{IS}_i$), we have $j \in \textit{IS}_i$.

According to Eq.~(\ref{eq:niv}) and Eq.~(\ref{eq:nov}), we indicate the number of window intervals in $\textit{IS}_i$ as $n_I(\textit{IS}_i)$, and the number of window positions in $\textit{IS}_i$ as $n_P(\textit{IS}_i)$. Note that the window intervals in $\textit{IS}_i$ are disjoint with each other. To facilitate the next ``interaction'' operation, we sort these intervals in ascending order, that is, $\textit{IS}_i[k].r < \textit{IS}_i[k+1].l$, where $\textit{IS}_i[k]$ is the $k^{th}$ window interval in $\textit{IS}_i$ (Line \ref{alg1:phase1:sort}).

\subsection{The Matching Algorithm}
Based on $\textit{IS}_i$ ($1\leq i\leq p$), we generate the final candidate set $\textit{CS}$ with an ``intersection'' operation. We first introduce the concept of \emph{candidate set} for $Q_i$, denoted as $\textit{CS}_i$ ($1\leq i\leq p$). For window $Q_1$, any window position $j$ in $\textit{IS}_1$ maps to a candidate subsequence $X(j,|Q|)$. Therefore, the candidate set for $Q_1$, denoted as $\textit{CS}_1$, is composed of all positions in $\textit{IS}_1$.  $\textit{CS}_1$ is still organized as a sequence of ordered non-overlapped window intervals, like $\textit{IS}_1$.

For $Q_2$, each  window position in $\textit{IS}_2$ also corresponds to a candidate subsequence. However, position $j$ in $\textit{IS}_2$ corresponds to the candidate subsequence $X(j-w,|Q|)$, because $X(j,w)$ is its second disjoint window. So the candidate set for $Q_2$, denoted as $\textit{CS}_2$, can be obtained by left-shifting each window position in $\textit{IS}_2$ with $w$. Similarly, $\textit{CS}_3$ is obtained by left-shifting the positions in $\textit{IS}_3$ with $2\cdot w$. In general, for window $Q_i$ ($1\leq i\leq p$), the candidate set $\textit{CS}_i$ is as follows,
\begin{equation}
\textit{CS}_i=\{j-(i-1)\cdot w|j\in \textit{IS}_i\}
\nonumber
\end{equation}

The shifting offset for $Q_i$ is denoted as $shift_i=(i-1)\cdot w$.
All candidate sets $\textit{CS}_i$ ($1 \leq i \leq p$) are still organized as an ordered sequence of non-overlapped window intervals. Moreover, it can be easily inferred that $n_I(\textit{CS}_i)=n_I(\textit{IS}_i)$ and $n_P(\textit{CS}_i)=n_P(\textit{IS}_i)$.

Through combining the lemmas in Section~\ref{sec:motivation} and the definition of $\textit{CS}_i$, we can obtain two important properties,
\begin{property}
    If $X(j,|Q|)$ is not contained by certain $\textit{CS}_i$ ($1\leq i\leq p$), then $X(j,|Q|)$ and $Q$ are not matched.
\label{pro:1}
\end{property}

\begin{property}
	If $X(j,|Q|)$ and $Q$ are matched,
	position $j$ belongs to all candidate sets $\textit{CS}_i$, that is,$j\in \textit{CS}_i \mbox{   } (1\leq i\leq p)$.
\end{property}

Now we present our approach to intersect $\textit{CS}_i$'s to generate the final $\textit{CS}$. It consists of $p$ rounds (Line \ref{alg1:phase1:begin}-\ref{alg1:phase1:end}). In the first round, we fetch $\textit{RList}_1$ from the index, and generate $\textit{IS}_1$ and $\textit{CS}_1$. We initialize $\textit{CS}$ as $\textit{CS}_1$. In the second round, we fetch $\textit{RList}_2$, and generate $\textit{CS}_2$ by shifting all window intervals in $\textit{IS}_2$ with $(2-1)\cdot w = w$ (Line \ref{alg1:phase1:shifting:begin}-\ref{alg1:phase1:shifting:end}). Then we intersect $\textit{CS}$ with $\textit{CS}_2$ to obtain up-to-date $\textit{CS}$ (Line \ref{alg1:phase1:intersection}). Because all intervals in $\textit{IS}_i$, as well as $\textit{CS}_i$, are ordered, the intersection operation can be executed by sequentially intersecting window intervals of $\textit{CS}$ and $\textit{CS}_2$, which is
quite similar to merge-sort algorithm with $O(n_I(\textit{CS})+n_I(\textit{CS}_2))$ complexity.
In general, during the $i$-th round, we intersect $\textit{CS}_i$ with $\textit{CS}$ of the last round, and generate the up-to-date $\textit{CS}$. After $p$ rounds, we obtain the final candidate set $\textit{CS}$.

%

\begin{figure}
	\centering
	\includegraphics[width=\linewidth]{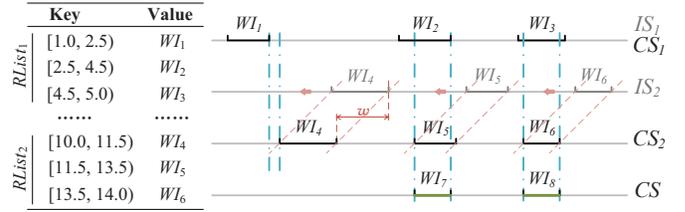}
	\caption{\label{fig:intersection}Example of the matching algorithm}
\end{figure}

We illustrate the algorithm with the example in Fig.~\ref{fig:intersection}. $\textit{RList}_1$ contains three intervals, $\textit{WI}_1$, $\textit{WI}_2$ and $\textit{WI}_3$. $\textit{RList}_2$ contains three intervals, $\textit{WI}_4$, $\textit{WI}_5$ and $\textit{WI}_6$. $\textit{IS}_1$ (or $\textit{IS}_2$) contains all the intervals covered by $\textit{RList}_1$ (or $\textit{RList}_2$). $\textit{CS}_1$ equals to $\textit{IS}_1$, while $\textit{CS}_2$ is generated by left-shifting $\textit{IS}_2$ with offset $w$. Then we intersect $\textit{CS}_1$ and $\textit{CS}_2$ to get $\textit{CS}$ in the second round, which is composed of $\textit{WI}_7$ and $\textit{WI}_8$.

In phase 2, according to $\textit{CS}$, we fetch data to generate the final qualified results (Line \ref{alg1:phase2:begin}-\ref{alg1:phase2:end}). Formally, for each window interval $\textit{WI}$ in $\textit{CS}$, we fetch the subsequences $X(\textit{WI}.l,\textit{WI}.r-\textit{WI}.l+|Q|)$ from data. Note that this subsequence contains $|\textit{WI}|$ number of subsequences. For each fetched length-$|Q|$ subsequence, we calculate the distance from $Q$ and return the qualified ones. If the query is cNSM query, each subsequence needs to be normalized before computing the ED or DTW distance. Moreover, most lower bounds used in UCR Suite~\cite{kdd12} can be also used here to speed up the verification, particularly for DTW measure.

%% file: tex/dynamic1.extended.tex
\section{KV-\MakeLowercase{match}$_{\text{DP}}$}
\label{sec:dynamic}

The basic KV-match uses a fixed window length $w$ to process the query, regardless of the query length. It has two limitations. First, the length of the supported query is limited. 
Second, we have less chance to exploit the characteristics of the query and the time series data to speed up processing. 

In this section, we propose KV-match$_{\text{DP}}$, which is based on multiple indexes with variable window lengths. Formally, the lengths of windows to build the index are summarized by two parameters, $w_u$ and $L$, where $w_u$ is the minimum window length and $L$ is the number of indexes. Then, the set of window lengths is $\Sigma=\{w_u * 2^{i-1} | 1 \leq i \leq L\}$. For example, suppose $w_u=25$ and $L=5$, we build indexes of length $25$, $50$, $100$, $200$ and  $400$ respectively. We use KV-index$_w$ to denote the index based on length-$w$ windows.
The set of indexes can be built simultaneously by extending the index building algorithm in Section~\ref{sec:index_building_algorithm} easily.

\subsection{Dynamic Query Segmentation}
We process the query with multiple indexes simultaneously. That is, we split $Q$ into a sequence of disjoint windows of variable lengths, $\{Q_1, Q_2,\cdots, Q_p\}$, and process each $Q_i$ with KV-index$_{|Q_i|}$, which is more flexible to utilize the characteristics of the data.
Once $Q$ is split, the following process is similar to that in KV-match. The only difference is that for window $Q_i$, we fetch $\textit{RList}_i$ from index KV-index$_{|Q_i|}$. Note that although in Lemmas in Section~\ref{sec:motivation}, $Q$ is split into equal-length windows, we can easily extend them to variable-length windows, since the proof always involves only one window.

The challenge here is how to split query $Q$ to achieve the best performance. We use \emph{query segmentation} to represent the result of query splitting. A segmentation, denoted as $\textit{SG}=\{r_1, r_2, \cdots, r_p\}$, means that $Q_1=Q(1,r_1)$, $Q_2=Q(r_1+1,r_2-r_1)$ and so on.
A high-quality segmentation should satisfy: 1) the length of each window belongs to $\Sigma$; 2) processing $Q$ with these windows results in high performance. 
We take the segmentation as an optimization problem and design an objective function to measure its quality.

\needspace{10mm}
\subsection{The Objective Function}
We first analyze the key factors to impact the efficiency. The runtime of query processing $T$ is composed of $T_1$ and $T_2$, those of phase 1 and 2 respectively. 
According to our theoretical analysis and experimental verification, $T_2$ is more significant to the efficiency, 
while $T_1$ is more stable. So we utilize the efficiency of phase 2 to measure the segmentation quality. Phase 2 consists of two parts, data fetching and distance computation, the former of which, determined by $n_I(\textit{CS})$, is much more time-consuming.

Therefore, for a segmentation $\textit{SG}$ of $Q$, after obtaining the final candidate set $\textit{CS}$, we use $n_I(\textit{CS})$ to measure the quality of $\textit{SG}$. The smaller $n_I(\textit{CS})$, the higher quality of $\textit{SG}$.
The challenge is we cannot obtain the exact value of $n_I(\textit{CS})$ without going through the index-probing phase. Moreover, although we can obtain the size of $n_I(\textit{CS}_i)$'s from the meta table, we cannot compute $n_I(\textit{CS})$ with $n_I(\textit{CS}_i)$'s directly.

To address this issue, we propose an objective function to estimate the value of $n_I(\textit{CS})$. The estimation is based on two assumptions. First, $\textit{IS}_i$'s of disjoint windows are independent with each other ($1\leq i \leq p$). Second, the size of each window interval in $\textit{IS}_i$ is much smaller than $|X|$. So we can take each window interval as a single point in $X$, and these positions are distributed uniformly.

Next, we introduce our objective function, denoted as $\mathcal{F}$. Assume that we use $\textit{SG}$ to split $Q$ into $Q_1, Q_2, \cdots, Q_p$, and obtain the size of each $\textit{IS}_i$ ($1\leq i\leq p$) based on the meta table. Then we estimate $n_I(\textit{CS})$ as follows. Based on these two assumptions, we can use $\frac{n_I(\textit{IS}_1)}{n}$ to approximately represent the probability of an interval contained in $\textit{CS}_1$, where $n$ is the length of $X$. It follows that $\frac{n_I(\textit{IS}_1)}{n} * \frac{n_I(\textit{IS}_2)}{n}$ is the probability of an interval contained in $\textit{CS}_1 \cap \textit{CS}_2$. Therefore, $\prod_{i=1}^{p}\frac{n_I(\textit{IS}_i)}{n}$ is the probability of an interval contained in the final $\textit{CS}$, which is proportional to $n_I(\textit{CS})$. It is obvious that the larger $p$, the smaller $\prod_{i=1}^{p}\frac{n_I(\textit{IS}_i)}{n}$. So, to eliminate the effect of number of windows, we take geometric mean of this value as the final objective function $\mathcal{F}$, as follows,

\small
\begin{equation}
	\mathcal{F}(\textit{SG}) = \sqrt[\leftroot{0}\uproot{25}p]{\prod_{i=1}^{p}\frac{n_I(\textit{IS}_i)}{n}} = \frac{1}{n}\sqrt[\leftroot{0}\uproot{25}p]{\prod_{i=1}^{p} n_I(\textit{IS}_i)}
	\label{eq:objection_function}
\end{equation}
\normalsize

The target segmentation is the one with the minimal value of $\mathcal{F}$\footnote{Since $\frac{1}{n}$ is a constant, we ignore it in the algorithm.}.

\subsection{Two-dimensional DP Approach}
We propose a two-dimensional dynamic programming algorithm to find the optimal $\textit{SG}$. We first define the search space. Since the length of each window $Q_i$ must belong to $\Sigma$, so in any $\textit{SG}=\{r_1, r_2,\cdots, r_p\}$, $r_i$ must be multiple times of $w_u$. Any $\textit{SG}$ not satisfying this constraint is invalid. Given query $Q=(q_1,q_2,\cdots,q_{m})$, we define the search space with sequence $Z=(1, 2,\cdots, m')$, where $m'=\lfloor\frac{m}{w_u}\rfloor$. Note that the values in $Z$ do not have impact on the generation of $\textit{SG}$. The only effect of $Z$ is to constrain the search space of $\textit{SG}$. Instead of finding $\textit{SG}$ on $Q$ directly, we find it from $Z$, denoted as $\textit{SG}_Z$, and then map it to $\textit{SG}$ of $Q$ by multiplying each endpoint of $Z$ with $w_u$. For example, let $|Q|=200$, $w_u=25$ and $L=3$. That is, we have three indexes, KV-index$_{25}$, KV-index$_{50}$ and KV-index$_{100}$. $\textit{SG}_Z=\{2, 6, 7, 8\}$ corresponds to $\textit{SG}=\{50, 150, 175, 200\}$. In this case, $Q$ is segmented into four windows, $Q(1,50)$, $Q(51,100)$, $Q(151,25)$ and $Q(176,25)$.

We search the optimal $\textit{SG}_Z$ with two-dimensional dynamic programming from left to right on $Z$ sequentially. The first dimension represents the boundaries of segmentation, and the second represents the number of windows contained in a segmentation. We use $v_{i,j}$ to represent a sub-state of calculation process, which corresponds to the best segmentation of the prefix of $Z$, $Z(1, i)$, with $j$ number of windows. For any $j$ ($1 \leq j \leq m'$), the best segmentation is the one with minimum $v_{m',j}$. After obtaining all $v_{m',j}$'s, we select the minimal one as the final $\textit{SG}_Z$, and map it to $\textit{SG}$.
The dynamic programming equation is presented as Eq.~(\ref{eq:dp}).

In Eq.~(\ref{eq:dp}), $\varphi$ represents the possible lengths of the window ending at $i$ in $\textit{SG}_Z$, and it has $L$ possible values at most. $C_{i-\varphi+1, \varphi}$ is the value of $n_I(\textit{IS})$ for the disjoint window $Q((i-\varphi)*w_u+1,\varphi *w_u)$, which can be obtained from the meta table of KV-index$_{\varphi *w_u}$, as explained in Section~\ref{sec:matching}.
The optimal $\textit{SG}_Z$ and $\textit{SG}$ can be recovered by leveraging backward-pointers.

\vspace{-4mm}
\small
\begin{equation}
	v_{i,j}= \left\{
	\begin{array}{l@{\;,\;}l}
	1 & i=0 \wedge j=0 \\
	+\infty & i = 0 \vee j = 0 \\
	\displaystyle\min_{\mathclap{\substack{\varphi=2^{k-1} \\ \mathmakebox[130\fboxrule][r]{1} \leq k \leq \min(L, \log_2(i)+1)}}}
		\sqrt[\leftroot{-3}\uproot{8}j]{(v_{i-\varphi, j-1})^{j-1} * C_{i-\varphi+1, \varphi}} & 1 \leq j \leq i \leq m'
	\end{array}
	\right.
	\label{eq:dp}
\end{equation}
\normalsize

\vspace{-1mm}
The complete algorithm is shown in Algorithm~\ref{alg:dp}.

\begin{algorithm}[t]
	\small
	\caption{Segment($w_u, L, Q$)}
	\begin{algorithmic}[1]
		\State $m' \gets \left\lfloor\frac{|Q|}{w_u}\right\rfloor, v_{i,j} \gets +\infty, P_{i,j} \gets -1$ $(0 \leq i \leq m')$ \label{alg3:initial:begin}
		\State $v_{0,0} \gets 1$ \label{alg3:initial:end}
		\For {$i \gets 1, m'$} \label{alg3:dp:begin}
		\For {$j \gets 1, i$}
		\For {$k \gets 1, \min(L, \log_2(i)+1)$}
		\State $\varphi \gets 2^{k-1}$
		\If {$\sqrt[\leftroot{-3}\uproot{8}j]{(v_{i-\varphi, j-1})^{j-1} * C_{i-\varphi+1, \varphi}} < v_{i,j}$}
		\State $v_{i,j} \gets \sqrt[\leftroot{-3}\uproot{8}j]{(v_{i-\varphi, j-1})^{j-1} * C_{i-\varphi+1, \varphi}}$
		\State $P_{i,j} \gets \varphi$
		\EndIf
		\EndFor
		\EndFor
		\EndFor \label{alg3:dp:end}
		\State $SG \gets \varnothing, i \gets m', j \gets \underset{x}{\arg\min}(v_{m', x})$ $(1 \leq x \leq m')$ \label{alg3:recover:begin}
		\While {$i \neq -1$}
		\State $SG$.add($i * w_u$)
		\State $i \gets i - P_{i,j}, j \gets j - 1$
		\EndWhile \label{alg3:recover:end}
		\State \Return $SG$	
	\end{algorithmic}
	\label{alg:dp}
\end{algorithm}

\emph{Analysis.} It happens that a large amount of windows of $X$ have similar mean values. In this case, certain rows in KV-index will have large value of $n_I$, which incurs large I/O cost to fetch $RList$ and large computation cost to merge $CS_i$ in each round. The KV-index$_{\text{DP}}$ can alleviate this phenomenon to some extent, since the objective function prefer the query windows with smaller $n_I$.

Moreover, we can use some techniques to alleviate this phenomenon further. First, to reduce the duplicate index visit, we can cache the index rows already fetched. Then for each new $RList$, if partial of it is already in the cache, we only need to fetch the rest part from KV-index. Second, we can reorder  $Q_i$'s to be processed according to the size of $RList_i$, which can be obtained easily from the meta data. In other words, we first process $Q_i$ with smaller $RList_i$, which can reduce both I/O cost and the merge computation cost. Third, note that each $CS_i$ is the \emph{superset} of the true result, so we can only process a partial of query windows, instead of all of them, to obtain the final $CS$ without loss of correctness. By combining the second and third optimization, we can skip some rows with large $n_I$ by ranking them at the bottom position.

%% file: tex/implementation.tex
\section{Implementation}
\label{sec:implementation}
We implement two versions of our approach to show the compatibility of our approach. One stores indexes in local disk files, and the other stores indexes on HBase~\cite{hbase}. Both are implemented with Java. The code and synthetic data generator are publicly available\footnote{https://github.com/DSM-fudan/KV-match}.

\subsection{Local File Version}
To compare the efficiency with previous subsequence matching methods, we first implement KV-match on conventional disk files.

In data file, all time series values are stored one by one in binary format, and their offsets are omitted because they can be easily inferred from bytes' length. In index file, the rows of KV-index are also stored contiguously. The offset of each row is recorded in meta data, stored at the footer of the file. The meta data will be retrieved first before processing the query. The start offset and length of each sequential read can be inferred by binary search on the meta data, and then a seek operation will be used to fetch data from file.

\subsection{HBase Table Version}
To verify the performance of KV-match for large data scale and test the scalability of our approach, we also implement it on HBase, where time series data and index are stored in tables respectively.

In time series table, time series is split into equal-length (1024 by default) disjoint windows, and each one is stored as a row. The key is the offset of the window, and value is the corresponding series data. In index table, a row of KV-index is stored as a row in HBase, and the meta table is also compacted to store as a row. We load the meta table to memory before processing the query. To take full advantage of the cluster, we adapt index building algorithm to the MapReduce framework.

\subsection{Compatibility with Other Systems}
Moreover, our index structure can be easily transplanted to other modern TSDB's. The only requirement is the system provides the ``scan'' operation to perform sequential data retrieval. Many systems support this operation, As examples, Table~\ref{tab:scan} lists the API used to implement the scan operation on some popular storage systems.

\begin{table}[htbp]
	\centering
	\caption{Scan Operation on Popular Storage Systems}
	\label{tab:scan}
	\begin{tabular}{@{\extracolsep{2pt}}ll@{}}
		\hline
		\textbf{System} & \textbf{Code Snippet of Retrieving Data in Specific Range} \\
		\hline
		\multirow{3}{*}{Local}  & \small\texttt{raf = new RandomAccessFile(file, "r");} \\
		& \small\texttt{raf.seek(offset);} \\
		& \small\texttt{raf.read(result, 0, length);} \\
		\hline
		\multirow{3}{*}{HDFS}  & \small\texttt{fdis = FileSystem.get(conf).open(path);} \\
		& \small\texttt{fdis.seek(offset);} \\
		& \small\texttt{fdis.read(result, 0, length);} \\
		\hline
		\multirow{2}{*}{HBase} & \small\texttt{scan = new Scan(startKey, endKey);} \\
		& \small\texttt{results = table.getScanner(scan);} \\
		\hline
		\multirow{3}{*}{LevelDB} & \small\texttt{for (it->Seek(startKey); it->Valid() \&\&}\\
		& \small\texttt{~~~~~it->key().ToString() < endKey;} \\
		& \small\texttt{~~~~~it->Next()) $\dots$} \\
		\hline
		\multirow{2}{*}{Cassandra} & \small\texttt{SELECT * FROM table WHERE} \\
		& \small\texttt{key >= startKey AND key < endKey} \\
		\hline
	\end{tabular}
\end{table} 

%% file: tex/experiment.extended.tex
\needspace{10mm}
\section{Experiments}
\label{sec:experiment}
In this section, we conduct extensive experiments to verify the effectiveness and efficiency of the proposed approach.

\subsection{Datasets and Settings}

\subsubsection{Real Datasets}
UCR Archive~\cite{UCRArchive} is a popular time series repository, which includes many datasets widely used in time series mining research. We concatenate the time series in UCR Archive to obtain desired length time series.

\subsubsection{Synthetic Datasets}
We use synthetic time series to test the scalability of our approach. The series are generated by combining three types of time series as follows.
\begin{itemize}
	\item Random walk. The start point and step length are picked randomly from $[-5, 5]$ and $[-1, 1]$ respectively;
	\item Gaussian. The values are picked from a Gaussian distribution with mean value and standard deviation randomly selected from $[-5, 5]$ and $[0, 2]$ respectively;
	\item Mixed sine. It is a mixture of several sine waves whose period, amplitude and mean value are randomly chosen from $[2, 10]$, $[2, 10]$ and $[-5, 5]$ respectively.
\end{itemize}

To generate a time series $X$, we execute the following steps repeatedly until $X$ is fully generated: i) randomly choose a type $t$, a length $l$ and the parameters according to type $t$; ii) generate a length-$l$ subsequence using type $t$ with parameters.

\subsubsection{Counterpart Approaches}
For RSM, we compare our approach (\textit{KVM} for short) with two index-based approaches, General Match~\cite{gmatch02} for ED and DMatch~\cite{Fu2008} for DTW. For cNSM, we compare with UCR Suite~\cite{kdd12} and FAST~\cite{edbt17}.

General Match~\cite{gmatch02} (\textit{GMatch} for short) is a classic R*-tree based approach for ED. We use the code from author, which
stores indexes in local disk files. Since building and updating R*-tree in distributed environment is not straightforward, we only compare it with our local file version. 

DMatch~\cite{Fu2008} is a duality-based subsequence matching approach for DTW, which is quite similar to other tree-style approaches. Because its code is not publicly available, we implement a C++ version based on General Match framework. The window length is set to 64 and each window is transformed to a 4-dimensional point by PAA. 

UCR Suite~\cite{kdd12} (\textit{UCR} for short) finds the best normalized matching subsequence under both ED and DTW. It scans the whole time series data, and uses some lower-bound techniques to speed up the query processing. Its code is publicly available\footnote{http://www.cs.ucr.edu/\raise.17ex\hbox{$\scriptstyle\sim$}eamonn/UCRsuite.html}, which is implemented in C++ and reads data on local disks. To make the comparison fair, we alter it to $\varepsilon$-match problem. Moreover, we implement a Java version to retrieve data on HBase, and conduct experiments for both local file and HBase table version to compare its scalability with KV-match.

FAST~\cite{edbt17} is a recent improvement on UCR Suite, which adds more lower-bound techniques to reduce the number of distance calculations. We use the code from author, and compare it with our local file version under both ED and DTW.

\subsubsection{Default Setting}
In KV-match$_{\text{DP}}$, $L$ is set to 5, and $\Sigma = \left\{25, 50, 100, 200, 400\right\}$. In index building algorithm, the initial fixed width $d$ is set to $0.5$ and the merge threshold $\gamma$ is set to $80\%$.
All experimental results are averaged over 100 runs.


To test the performance of processing queries with arbitrary lengths, we generate queries of length $128, 256, \cdots, 8192$. For each length, 100 different query series are generated. 


Experiments are executed on a cluster consisting of 8 nodes with HBase 1.1.5 (1 Master and 7 RegionServers). Each node is powered by Linux, and has two Intel Xeon E5 1.8GHz CPUs, 64GB memory, 5TB HDD storage. Experiments using local file version are executed on a single node of the cluster.

\begin{table}[t]
	\centering
	\caption{Results of RSM Queries Under ED Measure}
	\label{tab:exp1-ed}
	\begin{tabular}{@{\extracolsep{4pt}}ccrrr@{}}
		\hline
		\multirow{2}{*}{\textbf{Approach}} & \multirow{2}{*}{\textbf{Selectivity}} & \multirow{2}{*}{\textbf{\#candidates}} & \textbf{\#index} & \multirow{2}{*}{\textbf{Time} (ms)} \\
		& & & \textbf{accesses} & \\
		\hline
		& \textbf{$\bm{10^{-9}}$} &        13.9 & 279.2 &    852.3 \\
		&	\textbf{$\bm{10^{-8}}$} &      1837.5 & 240.1 &    541.2 \\
		\textbf{GMatch}   & \textbf{$\bm{10^{-7}}$} &   239,857.4 & 226.2 &  5,817.5 \\
		& \textbf{$\bm{10^{-6}}$} & 1,223,370.6 & 338.0 & 30,351.7 \\
		& \textbf{$\bm{10^{-5}}$} & 1,410,563.0 & 313.6 & 34,916.4 \\
		\hline
		& \textbf{$\bm{10^{-9}}$} &     2,754.9 &   4.6 &     60.4 \\
		&	\textbf{$\bm{10^{-8}}$} &     6,313.2 &   4.5 &     70.8 \\
		\textbf{KVM-DP}   & \textbf{$\bm{10^{-7}}$} &    29,853.1 &   4.4 &    138.8 \\
		& \textbf{$\bm{10^{-6}}$} &   113,434.1 &   6.0 &    567.4 \\
		& \textbf{$\bm{10^{-5}}$} &   153,565.1 &   7.0 &  1,200.7 \\
		\hline
	\end{tabular}
\end{table}

\begin{table}[t]
	\centering
	\caption{Results of RSM Queries Under DTW Measure}
	\label{tab:exp1-dtw}
	\begin{tabular}{@{\extracolsep{4pt}}ccrrr@{}}
		\hline
		\multirow{2}{*}{\textbf{Approach}} & \multirow{2}{*}{\textbf{Selectivity}} & \multirow{2}{*}{\textbf{\#candidates}} & \textbf{\#index} & \multirow{2}{*}{\textbf{Time} (ms)} \\
		& & & \textbf{accesses} & \\
		\hline
		& \textbf{$\bm{10^{-9}}$} & 1,176,639.8 & 250.0 &     543.5 \\
		&	\textbf{$\bm{10^{-8}}$} & 1,278,894.9 & 276.1 &   1,424.2 \\
		\textbf{DMatch}   & \textbf{$\bm{10^{-7}}$} & 1,800,014.9 & 447.8 &   7,847.2 \\
		& \textbf{$\bm{10^{-6}}$} & 2,406,697.3 & 619.2 &  29,952.9 \\
		& \textbf{$\bm{10^{-5}}$} & 3,431,349.8 & 902.9 & 132,062.4 \\
		\hline			
		& \textbf{$\bm{10^{-9}}$} &  25,423.9 &   4.7 &    115.3 \\
		&	\textbf{$\bm{10^{-8}}$} &  38,894.0 &   4.9 &    120.5 \\
		\textbf{KVM-DP}   & \textbf{$\bm{10^{-7}}$} &  87,002.5 &   5.3 &    634.1 \\
		& \textbf{$\bm{10^{-6}}$} & 118,580.9 &   6.6 &  3,641.3 \\
		& \textbf{$\bm{10^{-5}}$} & 218,965.5 &   7.1 & 21,348.2 \\
		\hline
	\end{tabular}
\end{table}
\begin{table}[t]
	\centering
	\caption{Results of cNSM Queries Under ED Measure}
	\label{tab:exp2-ed}
	\begin{tabular}{@{\extracolsep{3.5pt}}clrrrrr@{}}
		\hline
		\multirow{2}{*}{\textbf{Selectivity}} & \multicolumn{4}{c}{\textbf{KVM-DP} (s)} & \multicolumn{1}{c}{\textbf{UCR}} & \multicolumn{1}{c}{\textbf{FAST}} \\
		\cline{2-5}\cline{6-6}\cline{7-7}
		& \multicolumn{1}{c}{$\bm{\alpha}$\textbackslash $\bm{\beta'}$} & \multicolumn{1}{c}{\textbf{1.0}} & \multicolumn{1}{c}{\textbf{5.0}} & \multicolumn{1}{c}{\textbf{10.0}} & \multicolumn{1}{c}{\textbf{Avg.(s)}} & \multicolumn{1}{c}{\textbf{Avg.(s)}}\\
		\hline
		\multirow{3}{*}{$\bm{10^{-9}}$}  & \textbf{1.1} & 0.51 & 2.33 & 4.64 & \multirow{3}{*}{59.84} & \multirow{3}{*}{86.05}\\ 
		& \textbf{1.5} & 0.56 & 2.58 & 5.05 &\\ 
		& \textbf{2.0} & 0.59 & 2.70 & 5.51 &\\
		\hline
		\multirow{3}{*}{$\bm{10^{-8}}$}  & \textbf{1.1} & 0.72 & 3.22 & 6.18 & \multirow{3}{*}{60.17} & \multirow{3}{*}{86.09}\\ 
		& \textbf{1.5} & 1.00 & 4.60 & 8.98 &\\ 
		& \textbf{2.0} & 1.22 & 5.47 & 10.66 &\\
		\hline
		\multirow{3}{*}{$\bm{10^{-7}}$} & \textbf{1.1} & 1.30 & 5.46 & 10.29 & \multirow{3}{*}{65.25} & \multirow{3}{*}{87.79}\\ 
		& \textbf{1.5} & 2.82 & 11.53 & 21.75 &\\ 
		& \textbf{2.0} & 3.72 & 16.20 & 29.15 &\\
		\hline
		\multirow{3}{*}{$\bm{10^{-6}}$} & \textbf{1.1} & 1.69 & 6.74 & 14.53 & \multirow{3}{*}{69.17} & \multirow{3}{*}{88.64}\\ 
		& \textbf{1.5} & 3.15 & 15.19 & 27.53 &\\ 
		& \textbf{2.0} & 4.39 & 20.77 & 35.75 &\\
		\hline
		\multirow{3}{*}{$\bm{10^{-5}}$} & \textbf{1.1} & 1.94 & 7.82 & 12.92 & \multirow{3}{*}{70.59} & \multirow{3}{*}{89.83}\\ 
		& \textbf{1.5} & 4.23 & 15.98 & 28.26 &\\ 
		& \textbf{2.0} & 5.77 & 21.55 & 37.66 &\\
		\hline
	\end{tabular}
\end{table}

\begin{table}[t]
	\centering
	\caption{Results of cNSM Queries Under DTW Measure}
	\label{tab:exp2-dtw}
	\begin{tabular}{@{\extracolsep{1.5pt}}clrrrrr@{}}
		\hline
		\multirow{2}{*}{\textbf{Selectivity}} & \multicolumn{4}{c}{\textbf{KVM-DP} (s)} & \multicolumn{1}{c}{\textbf{UCR}} & \multicolumn{1}{c}{\textbf{FAST}}  \\
		\cline{2-5}\cline{6-6}\cline{7-7}
		& \multicolumn{1}{c}{$\bm{\alpha}$\textbackslash $\bm{\beta'}$} & \multicolumn{1}{c}{\textbf{1.0}} & \multicolumn{1}{c}{\textbf{5.0}} & \multicolumn{1}{c}{\textbf{10.0}} & \multicolumn{1}{c}{\textbf{Avg.(s)}} & \multicolumn{1}{c}{\textbf{Avg.(s)}} \\
		\hline
		\multirow{3}{*}{$\bm{10^{-9}}$}  & \textbf{1.1} & 0.72 & 2.71 & 3.71 & \multirow{3}{*}{139.57} & \multirow{3}{*}{77.5}\\ 
		& \textbf{1.5} & 0.66 & 2.97 & 4.72 &\\ 
		& \textbf{2.0} & 0.78 & 3.37 & 6.00 &\\
		\hline
		\multirow{3}{*}{$\bm{10^{-8}}$} & \textbf{1.1} & 0.89 & 2.66 & 5.31 & \multirow{3}{*}{140.06} & \multirow{3}{*}{78.57}\\ 
		& \textbf{1.5} & 1.24 & 4.89 & 7.89 &\\ 
		& \textbf{2.0} & 1.43 & 5.01 & 9.21 &\\
		\hline
		\multirow{3}{*}{$\bm{10^{-7}}$} & \textbf{1.1} & 1.88 & 6.61 & 10.02 & \multirow{3}{*}{142.99} & \multirow{3}{*}{85.07}\\ 
		& \textbf{1.5} & 3.81 & 13.79 & 23.30 &\\ 
		& \textbf{2.0} & 4.46 & 15.92 & 33.00 & \\
		\hline
		\multirow{3}{*}{$\bm{10^{-6}}$} & \textbf{1.1} & 5.58 & 14.29 & 18.69 & \multirow{3}{*}{153.88} & \multirow{3}{*}{103.60}\\ 
		& \textbf{1.5} & 11.09 & 30.74 & 60.27 &\\ 
		& \textbf{2.0} & 11.40 & 33.72 & 60.56 &\\
		\hline
		\multirow{3}{*}{$\bm{10^{-5}}$} & \textbf{1.1} & 19.75 & 36.61 & 49.94 & \multirow{3}{*}{177.28} & \multirow{3}{*}{137.01}\\ 
		& \textbf{1.5} & 40.35 & 57.90 & 102.72 &\\ 
		& \textbf{2.0} & 44.07 & 76.23 & 106.97 &\\
		\hline
	\end{tabular}
\end{table}

\subsection{Results of RSM Queries}
\label{sec:experiment_gmatch}
We first compare KV-match$_{\text{DP}}$ with General Match and DMatch. The experiment is conducted on length-$10^9$ real dataset with queries of different selectivities. The results are shown in Table~\ref{tab:exp1-ed} and \ref{tab:exp1-dtw} respectively. 

It can be seen that when the selectivity increases, the number of candidates of General Match explodes dramatically, and in the case of higher selectivities, it is much larger than that of ours. Although General Match converts all values in a window into a multi-dimensional point, which keeps more information than the mean value used in KV-index, it generates candidates only based on one single window. In contrast, our approach combines the pruning power of multiple windows, which can achieve smaller candidate set. 

The number of index accesses of General Match is 20-30 times larger than that of ours. Due to fewer index accesses and less number of candidates, our approach achieves the overall performance improvement of one order of magnitude compared to General Match. An interesting phenomenon is that for queries of low selectivities ($10^{-8}$ or $10^{-9}$), the number of candidates of our approach is slightly larger than that of General Match. However, benefiting from fewer index accesses, we still achieve better overall performance.

Similar to General Match, DMatch also conducts large number of index accesses, and has to verify one or two orders of magnitude more candidates than ours. The reason is still the single window candidate generation mechanism and tree-style index structure, as General Match. 

\subsection{Influence of Window Size $w$}
In this experiment, we investigates the pruning performance of building index with the mean values. We compare the number of candidates obtained from each query window $Q_i$ of KV-match and FRM~\cite{frm94}~\footnote{FRM is a special case of General Match when $J=1$.}. FRM is selected to compare because its mechanism is analogous to KV-match. FRM builds the index based on the sliding windows of $X$, and each window is transformed into an $f$-dimensional point. Then the transformed points are stored in R-tree. To process query $Q$, FRM splits $Q$ into $p$ number of disjoint windows $Q_i$ ($1\leq i\leq p$). For each window, a set of candidates are obtained by a range query to R-tree. Then, the \emph{union} of candidates of all windows forms the final candidate set. In contrast, in KV-match, the final candidate sets, $CS$, is the intersection of $CS_i$'s.

In Table~\ref{tab:win-total-candidates}, we show the ratio of number of candidates per window between our approach and FRM. The experiments are conducted on time series of length $10^9$. We run queries of different selectivities. For each selectivity, 100 randomly generated queries of length 2048 are processed, and the number of candidates are averaged. We compare KV-indexes and FRM with variable window sizes, 50, 100, 200, 400. Moreover, we also show the ratio of the number of final candidates between our approach and FRM.

It can be seen that our approach will generate more candidates per window, $CS_i$, especially for smaller $w$ and larger $|Q|$, since the range depends on $\frac{\varepsilon}{w}$. However, the number of final candidates, $CS$, of our approach is much smaller than that of FRM, because in KV-match, $CS$ is the intersection of $CS_i$'s, while in FRM, $CS$ is the union of $CS_i$'s. Consider it is more expensive to fetch the time series to compute the distance, reducing $CS$ is more beneficial. Moreover, for each $Q_i$, we only visit index with a sequential scan operation, while in FRM we need to visit multiple index nodes, which may incur more I/O cost. Finally, the mechanism of KV-match${_{\text{DP}}}$ can avoid to use the query windows with many candidates.

\begin{table*}[htbp]
	\centering
	\caption{The Ratio of KV-match and FRM on Window Averaged Candidates vs. Final Total Candidates}
	\label{tab:win-total-candidates}
	\begin{tabular*}{\linewidth}{@{\extracolsep{\fill}}clrrrrrrrr@{}}
		\hline
		\multirow{2}{*}{\textbf{Selectivity}} & \multirow{2}{*}{$\bm{|Q|}$} & \multicolumn{4}{c}{\textbf{\#candidates per window}} & \multicolumn{4}{c}{\textbf{\#candidates in final}} \\
		\cline{3-6}\cline{7-10}
		&  & \multicolumn{1}{c}{$\bm{w=}$ \textbf{50}} & \multicolumn{1}{c}{\textbf{100}} & \multicolumn{1}{c}{\textbf{200}} & \multicolumn{1}{c}{\textbf{400}} & \multicolumn{1}{c}{$\bm{w=}$ \textbf{50}} & \multicolumn{1}{c}{\textbf{100}} & \multicolumn{1}{c}{\textbf{200}} & \multicolumn{1}{c}{\textbf{400}} \\
		\hline
		\multirow{5}{*}{$\bm{10^{-6}}$}	& \textbf{512} &  14.3 &  21.8 &  29.7 & 31.3 & 0.002 & 0.104 & 2.626 & 31.287 \\
		& \textbf{1024} &  40.5 &  58.7 & 47.9 & 20.8 & 0.081 & 0.086 & 0.750 &  7.055 \\
		& \textbf{2048} &  52.1 &  65.5 & 59.3 & 21.2 & 0.010 & 0.007 & 0.041 &   0.323 \\
		& \textbf{4096} &  65.5 &  69.8 & 64.4 & 37.9 & 0.112 & 0.040 & 0.029 &   0.143 \\
		& \textbf{8192} &  91.9 &  82.6 & 70.6 & 57.4 & 0.108 & 0.080 & 0.049 &   0.069 \\
		\hline			
		\multirow{5}{*}{$\bm{10^{-5}}$}	& \textbf{512} &  12.4 &   8.1 &   5.9 &   8.4 & 0.091 & 0.226 & 1.561 &   8.352 \\
		& \textbf{1024} &  18.3 &  10.1 &   7.0 &   5.8 & 0.184 & 0.029 & 0.062 &   1.044 \\
		& \textbf{2048} &  41.0 &  18.4 &  10.0 &  10.2 & 0.209 & 0.076 & 0.002 &   0.040 \\
		& \textbf{4096} &  81.1 &  33.6 &  18.2 &  15.6 & 0.247 & 0.131 & 0.025 &   0.006 \\
		& \textbf{8192} & 168.7 &  69.9 &  33.9 &  24.4 & 0.354 & 0.170 & 0.043 &   0.002 \\
		\hline			
		\multirow{5}{*}{$\bm{10^{-4}}$}	& \textbf{512} &  13.1 &   7.7 &   4.7 &   4.7 & 0.183 & 0.273 & 1.138 &   4.714 \\
		& \textbf{1024} &  23.7 &  10.3 &   5.5 &   3.4 & 0.204 & 0.029 & 0.080 &   0.587 \\
		& \textbf{2048} &  62.3 &  23.0 &   9.6 &   5.7 & 0.483 & 0.181 & 0.026 &   0.071 \\
		& \textbf{4096} & 165.0 &  60.3 &  24.5 &  11.3 & 0.752 & 0.582 & 0.388 &   0.137 \\
		& \textbf{8192} & 281.4 & 103.5 &  40.2 &  17.5 & 0.535 & 0.400 & 0.196 &   0.042 \\
		\hline		
		\multirow{5}{*}{$\bm{10^{-3}}$}	& \textbf{512} &  13.5 &   5.8 &   2.7 &   2.3 & 0.149 & 0.207 & 0.577 &   2.315 \\
		& \textbf{1024} &  28.9 &  11.6 &   5.5 &   2.6 & 0.340 & 0.099 & 0.171 &   0.553 \\
		& \textbf{2048} &  68.5 &  26.1 &  10.7 &   5.2 & 0.531 & 0.319 & 0.087 &   0.152 \\
		& \textbf{4096} & 161.8 &  61.4 &  24.3 &  10.0 & 0.728 & 0.520 & 0.280 &   0.063 \\
		& \textbf{8192} & 266.2 & 152.6 &  61.3 &  24.9 & 0.940 & 0.704 & 0.508 &   0.277 \\
		\hline
	\end{tabular*}
\end{table*}

\begin{table}[t]
	\centering
	\caption{Influence of $w$ on Index Size and Building Time}
	\label{tab:w-index}
	\begin{tabular*}{0.6\linewidth}{@{\extracolsep{\fill}}crr@{}}
		\hline
		$\bm w$ & \textbf{Size (MB)} & \textbf{Building time (s)} \\
		\hline
		\textbf{25}	 & 354.09 & 299.38 \\
		\textbf{50}	 & 287.21 & 234.30 \\
		\textbf{100} & 236.49 & 227.06 \\
		\textbf{200} & 194.52 & 210.18 \\
		\textbf{400} & 155.47 & 198.12 \\
		\hline
	\end{tabular*}
\end{table}

\subsection{Results of cNSM Queries}
In this experiment, we compare KV-match$_{\text{DP}}$ with UCR Suite and FAST for cNSM on local disk. The experiment is conducted on length-$10^9$ real dataset with queries of different selectivities. The results under ED and DTW measures are shown in Table~\ref{tab:exp2-ed} and \ref{tab:exp2-dtw} respectively. For each selectivity, we report the runtime for different $\alpha$ and $\beta$. The constraints are also embedded into UCR Suite and FAST, so unqualified candidates are abandoned too. For simplicity, we only report the average runtime for each selectivity, because theirs runtime for queries in the same selectivity group is quite similar.

We use relative offset shifting $\beta'$ in cNSM experiments, which is the percentage of the value range of the whole data series. Therefore, $\beta=(\max(X)-\min(X))* \beta'\%$.

It can be seen that when the selectivity increases, the runtime of KV-match increases steadily. When the selectivity is fixed, the runtime increases as $\alpha$ and $\beta$ increase. Because UCR Suite almost always scans the whole dataset, its runtime is more stable and dominated by I/O cost. The extra lower-bounds in FAST seems not efficient for ED, due to its overhead of data preparation. While for DTW, FAST achieves obvious improvement comparing to UCR Suite, especially for queries of low selectivities ($10^{-8}$ or $10^{-9}$). In most cases, our approach achieves the performance improvement of one to two orders of magnitude compared to them. 


\subsection{Index Size and Building Time}
\label{sec:experiment_size}
We compare the index space cost and building time of KV-match$_{\text{DP}}$ and DMatch. GMatch has similar space cost and building time as those of DMatch, and so we do not show them in the results. The experiment is conducted on the local file version with real datasets. Results are shown in Fig.~\ref{fig:exp6}.
We also show the size of time series data as dark blue bars.

It can be seen that the index sizes of both DMatch and KV-match$_{\text{DP}}$ are about 10\% of data size, and the size of KV-match$_{\text{DP}}$ is slightly larger than that of DMatch. However, KV-match$_{\text{DP}}$ consists of 5 KV-indexes, so the size of a single KV-index is much smaller than that of DMatch. We also show the index building time as lines in Fig.~\ref{fig:exp6}. Our index is much more efficient to build, due to its simple structure. In the extremely large data scale (the trillion-length time series), it takes 36 hours to build all 5 KV-indexes for KV-match$_{\text{DP}}$ on HBase.

Moreover, we test the influence of window size $w$ on the index size and building time. In Table~\ref{tab:w-index}, we show the index size and building time of KV-index with fixed $w$ on time series of length $10^9$. It can be seen that as $w$ increases, both index size and building time decrease gradually. This is because that larger $w$ makes the mean values of the adjacent windows more similar, and correspondingly makes $n_I(V_i)$ smaller, which reduces both the index size and the building time.


\subsection{Scalability}
\begin{figure}[t]
	\begin{minipage}[t]{0.49\linewidth}
		\includegraphics[width=40mm,height=30mm]{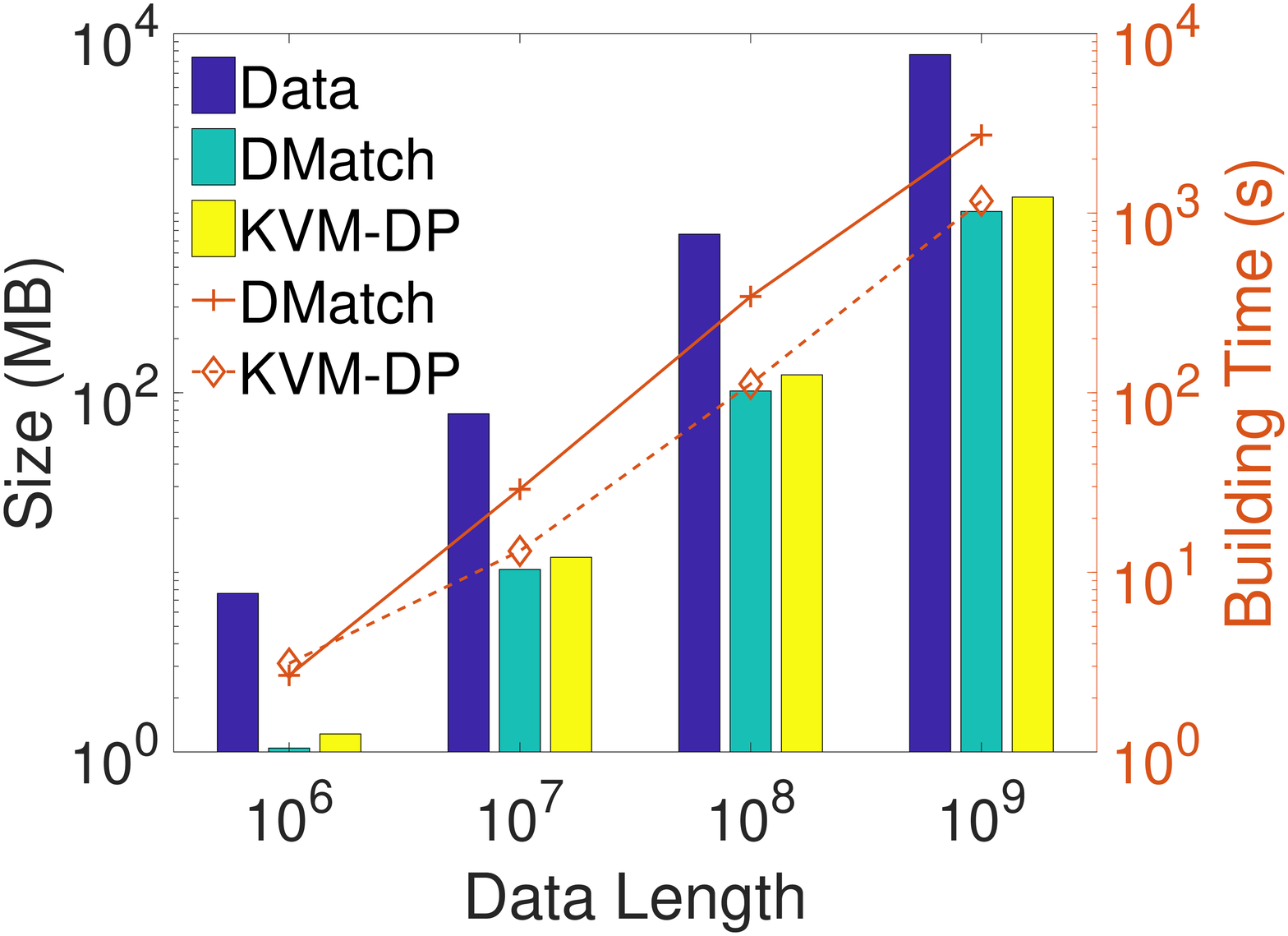}
		\caption{Size \& building time}
		\label{fig:exp6}
	\end{minipage}
	\hfill
	\begin{minipage}[t]{0.49\linewidth}
		\includegraphics[width=40mm,height=30mm]{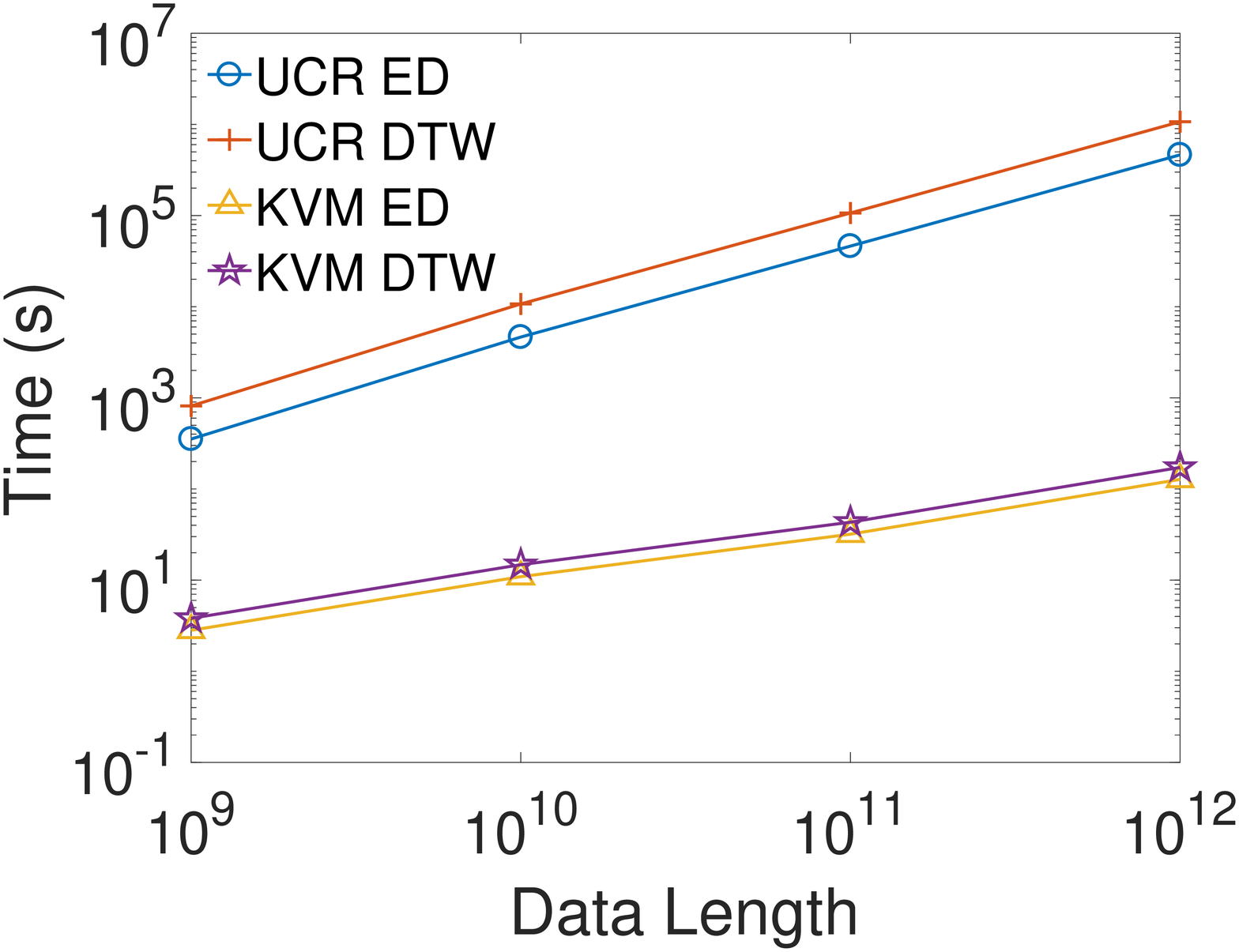}
		\caption{Scalability}
		\label{fig:exp2}
\end{minipage}
\end{figure}

To investigate the scalability of our approach, we use longer synthetic time series, from length-$10^9$ to length-$10^{12}$, to compare KV-match$_{\text{DP}}$ and UCR Suite for cNSM queries. Both time series data and our index is stored as HBase table, and both ED and DTW measures are compared. We set $\alpha=1.5$, $\beta'=1.0$, and hold selectivity to $10^{-7}$ by adjusting $\varepsilon$. The results are shown in Fig.~\ref{fig:exp2}.

\begin{figure}[t]
	\centering
	\begin{tabular}[htbp]{cc}
		\includegraphics[width=40mm,height=30mm]{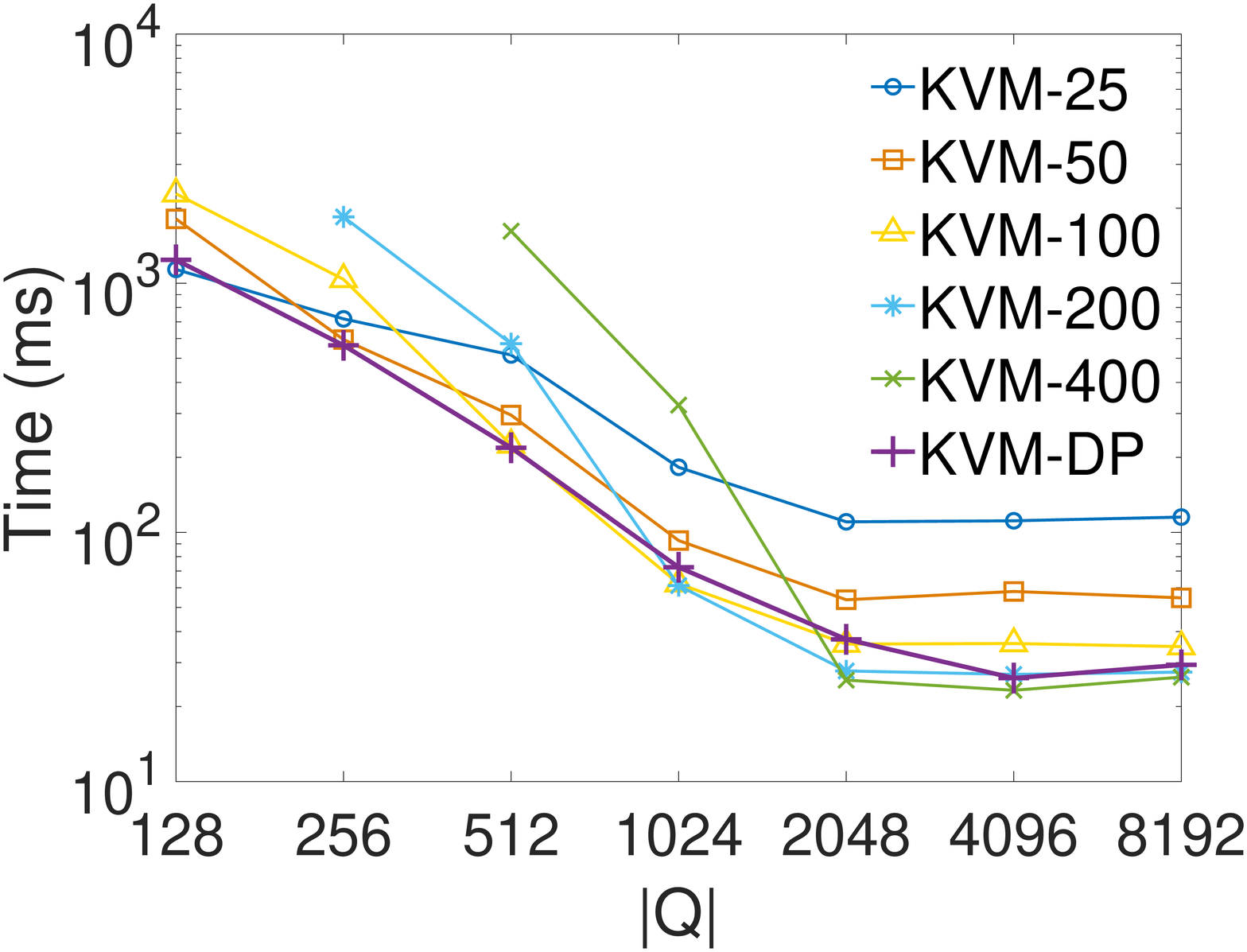} &
		\includegraphics[width=40mm,height=30mm]{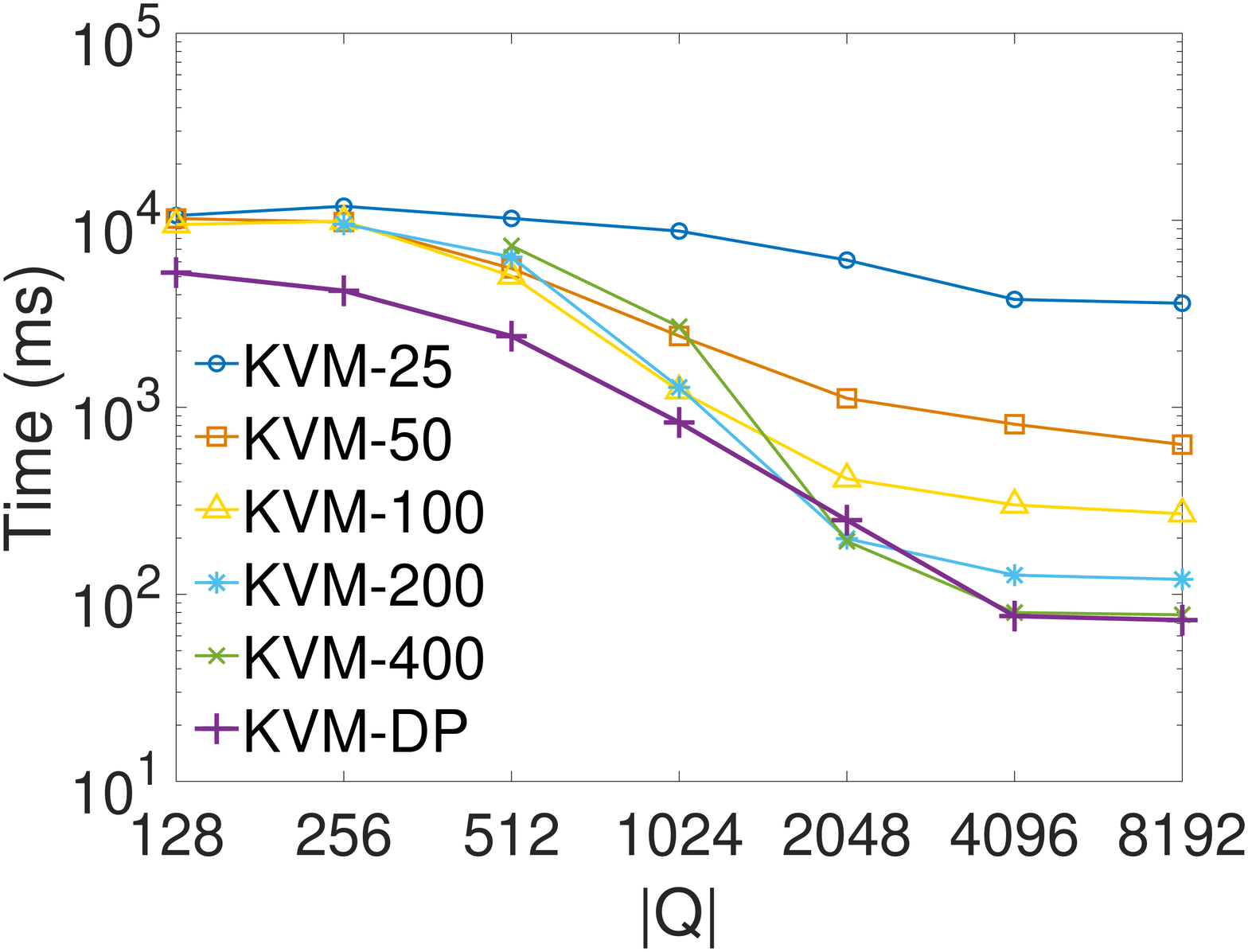} \\
		\footnotesize (a) $\varepsilon = 10$ & \footnotesize (b) $\varepsilon = 100$ \\
	\end{tabular}
	\caption{\label{fig:exp7}Effect of dynamic window segmentation}
\end{figure}

It can be seen that KV-match${_{\text{DP}}}$ is faster than UCR Suite under both ED and DTW measures by almost two to three orders of magnitude. For trillion-length ($10^{12}$) series, we can process queries by 127s (under ED measure) and 243s (under DTW measure) on average, which shows great scalability.

\subsection{KV-match$_{\textit{DP}}$ vs. the Basic KV-match}
\label{sec:experiment_dp}
In this experiment, we compare the runtime between KV-match$_{\text{DP}}$ and KV-match for RSM queries.
We build 5 KV-indexes with $w$ as $25, 50, 100, 200, 400$ respectively.
For KV-match$_{\text{DP}}$, we set $\Sigma=\left\{25,50,100,200,400\right\}$ to use all these indexes.
The experiment is conducted with local file version on length-$10^9$ real dataset. Because the performance of a single index is highly related to the length of queries, we test the runtime of variable query lengths. Fig.~\ref{fig:exp7} (a) and (b) show the results in the case of $\varepsilon=10$ (representing low selectivity) and $\varepsilon=100$ (representing high selectivity) respectively.

It can be seen that in most cases, KV-match$_{\text{DP}}$ outperforms all single indexes. On the contrary, the index with small window length is suitable only for shorter queries, while the index with large window length only works well on longer queries. The results verify the effectiveness of our query segmentation algorithm. KV-match$_{\text{DP}}$ can utilize the pruning power of multiple window lengths and leverage the data characteristics of the query sequences.

%% file: tex/related1.tex
\section{Related Work}
\label{sec:related}

Subsequence matching problem has been studied extensively in last two decades.

\textbf{Approaches for RSM problem.} The pioneering work~\cite{frm94}, FRM, used Euclidean distance as the similarity measure. It transforms each sliding window into a low-dimensional point and stores in R-tree. Disjoint windows of query series are also transformed and the candidates are retrieved by range queries on R-tree. To improve the efficiency, Dual-Match~\cite{dualmatch01} extracts disjoint windows from data series and sliding windows from query series, which reduces the size of R-tree. General Match~\cite{gmatch02} generalizes both of them, and benefits from both point filtering effect in Dual-Match and window size effect in FRM. \cite{multi07} builds multiple indexes and picks the optimal one to process the query according to the query length. All these approaches transform subsequences into low-dimensional points, and build R-tree as the index. This mechanism incurs large amount of index visits for large data scale. In contrast, KV-match only needs a scan operation for each $Q_i$.

Some works deal with RSM problem with other distance functions. The Dynamic Time Warping (DTW) distance is studied in \cite{lb_keogh}, which proposes two lower bounds for DTW, LB\_Keogh and LB\_PAA. Also, DMatch~\cite{Fu2008} presents a duality-based approach for DTW by extending Dual-Match~\cite{dualmatch01}.
\cite{vldb12} supports multiple distances which satisfy specific property. 

GDTW~\cite{icde18} is a general framework to apply the idea of DTW to more point-to-point distance functions. It is orthogonal to KV-match, because it focuses on the distance function while KV-match considers how to support both RSM and NSM queries simultaneously. Recently, adaptive approach is studied in whole matching problem~\cite{adaptive}, which first builds a coarse-granularity index, then refines it during the query processing. This mechanism can reduce the initial construction time, and also make the index evolved according to the queries. This work deals with the whole matching problem. It is not trivial to adapt it to support both RSM and NSM problem.

Although there exist some works to support both ED and DTW measure, all these works don't support normalization.

\textbf{Approaches for NSM problem.} In \cite{kdd12}, authors claim that normalization is vital and propose the UCR Suite to deal with normalized subsequence matching under both ED and DTW. Some optimizations are utilized to speed up. However, it needs to scan the whole sequence to find the qualifying subsequences, which is intolerable for large data scale. Recently, FAST~\cite{edbt17} is proposed to improve the efficiency. It is based on UCR Suite, and adds some lower-bound techniques to reduce the number of candidate verification. Similar with UCR suite, FAST still needs to scan the whole time sequence. In contrast, KV-match proposes an index to deal with cNSM problem, which is more efficient.  ONEX~\cite{vldb16} utilizes the marriage of ED and DTW to support the normalized subsequence search. It builds the index for all possible subsequence lengths. For each subsequence length, it first normalizes all subsequences, and then builds the index based on a clustering approach. So it cannot support RSM and NSM problems simultaneously.

In sum, only UCR Suite~\cite{kdd12} and FAST~\cite{edbt17} support both RSM and NSM~\footnote{Although they aim to process the NSM query, we can easily adapt them to deal with the RSM query by removing the normalization step.}. However, they need to scan the full time series. There is no existing work to build the index supporting both RSM and NSM problem.

%% file: tex/conclusion.tex
\section{Conclusion and Future Work}
\label{sec:conclusion}
We propose a novel constrained normalized subsequence matching problem (cNSM), which provides a knob to flexibly control the degree of offset shifting and amplitude scaling.
We also propose a key-value index structure KV-index, corresponding matching algorithm KV-match, and the extended version KV-match$_{\text{DP}}$, to support both RSM and cNSM problems under either ED or DTW measure.
Experimental results verify the efficiency and effectiveness.
To the best of our knowledge, this is the first index-based work for normalized subsequence matching. In the future, we will try to support more distance measures, especially variable-length DTW. 

%% file: tex/appendices.tex
\appendices
\section{Proof of Lemma~\ref{lemma:RSM-DTW}}
\label{app:1}
\small
By combining Eq.~(\ref{eq:lb_paa}) and $\textit{DTW}_\rho (S,Q)\leq \varepsilon$, we can easily infer the following three cases of $\mu_i^S$,
\begin{itemize}
	\item[(a)] $\mu_i^S > \mu_i^U$. In order to let $w\cdot (\mu_i^S-\mu_i^U)^2\leq \varepsilon$, $\mu_i^{S}$ should satisfy $\mu_i^U < \mu_i^S \leq \mu_i^U + \frac{\varepsilon}{\sqrt{w}}$;
	\item[(b)] $\mu_i^S < \mu_i^L$. In order to let $w\cdot (\mu_i^S-\mu_i^L)^2\leq \varepsilon$, $\mu_i^{S}$ should satisfy $\mu_i^L - \frac{\varepsilon}{\sqrt{w}} \leq \mu_i^S < \mu_i^L$;
	\item[(c)] Otherwise. Because $0\leq \varepsilon$ always holds, $\mu_i^L \leq \mu_i^S \leq \mu_i^U$.
\end{itemize}

Taking the union of above three cases, we will get Eq.~(\ref{eq:RSM-DTW}).
\qed

\section{Proof of Lemma~\ref{lemma:cNSM-DTW}}
\label{app:2}

Let $L'=\left(\frac{l_1-\mu_Q}{\sigma_Q},\cdots,\frac{l_m-\mu_Q}{\sigma_Q}\right)$,  $U'=\left(\frac{u_1-\mu_Q}{\sigma_Q},\cdots,\frac{u_m-\mu_Q}{\sigma_Q}\right)$
be two length-$m$ series derived from $L$ and $U$. Since $L'$ and $U'$ are derived by a simple linear transformation, it can be easily inferred that $L'$ and $U'$ are still the lower and upper envelop of $\hat{Q}=\left(\frac{q_1-\mu_Q}{\sigma_Q},\cdots,\frac{q_m-\mu_Q}{\sigma_Q}\right)$.

Similar to Lemma~\ref{lemma:RSM-DTW}, if $\textit{DTW}_{\rho}(\hat S,\hat Q)\leq \varepsilon$, we have
$\hat\mu^{S}_{i} \in \left[\mu_i^{L'} - \frac{\varepsilon}{\sqrt{w}}, \mu_i^{U'} + \frac{\varepsilon}{\sqrt{w}}\right]$,
where $\hat\mu^{S}_{i}$ is the mean value of the $i$-th windows of $\hat{S}$, $\mu_i^{L'}$ and $\mu_i^{U'}$ are the mean values of the $i$-th windows of $L'$ and $U'$ respectively.

By simple transformation, we have $\hat\mu^{S}_{i}=\frac{\mu_i^S-\mu^S}{\sigma^S}$, $\mu_i^{L'}=\frac{\mu_i^{L}-\mu^Q}{\sigma^Q}$ and $\mu_i^{U'}=\frac{\mu_i^{U}-\mu^Q}{\sigma^Q}$, so
\vspace{-2mm}
\footnotesize
\begin{equation}
\frac{\mu_i^S-\mu^S}{\sigma^S} \in \left[\frac{\mu_i^{L}-\mu^Q}{\sigma^Q} - \frac{\varepsilon}{\sqrt{w}}, \frac{\mu_i^{U}-\mu^Q}{\sigma^Q} + \frac{\varepsilon}{\sqrt{w}}\right]
\label{eq:lu-dtw}
\end{equation}\vspace{-2mm}
\small

In Eq.~(\ref{eq:lu-dtw}), $\mu_i^{L}$ and $\mu_i^{U}$ are the mean values of the $i$-th windows of $L$ and $U$ respectively.

Let $a=\frac{\sigma^{S}}{\sigma^{Q}}$ and $b=\mu^S-\mu^Q$,
By replacing $\sigma^S=a\sigma^Q$ and $\mu^S=\mu^Q+b$ in Eq.~(\ref{eq:lu-dtw}), we can get
\footnotesize
\begin{equation}
\mu_i^S \in \bigg[\left(\mu_i^L-\mu^Q-\frac{\varepsilon  \sigma^Q}{\sqrt{w}}\right) a + b + \mu^Q, \left(\mu_i^U-\mu^Q+\frac{\varepsilon  \sigma^Q}{\sqrt{w}}\right) a + b + \mu^Q\bigg]
\nonumber
\end{equation}
\small
where $a \in [\frac{1}{\alpha}, \alpha]$ and $b \in [-\beta,\beta]$. Similar to the proof of Lemma~\ref{lemma:cNSM-ED}, we can obtain that the range of $\mu_i^S$ is exactly Eq.~(\ref{eq:lemma4}).
\qed

%% file: kv-match.extended.bbl
\begin{thebibliography}{10}
	\providecommand{\url}[1]{#1}
	\csname url@samestyle\endcsname
	\providecommand{\newblock}{\relax}
	\providecommand{\bibinfo}[2]{#2}
	\providecommand{\BIBentrySTDinterwordspacing}{\spaceskip=0pt\relax}
	\providecommand{\BIBentryALTinterwordstretchfactor}{4}
	\providecommand{\BIBentryALTinterwordspacing}{\spaceskip=\fontdimen2\font plus
		\BIBentryALTinterwordstretchfactor\fontdimen3\font minus
		\fontdimen4\font\relax}
	\providecommand{\BIBforeignlanguage}[2]{{%
			\expandafter\ifx\csname l@#1\endcsname\relax
			\typeout{** WARNING: IEEEtran.bst: No hyphenation pattern has been}%
			\typeout{** loaded for the language `#1'. Using the pattern for}%
			\typeout{** the default language instead.}%
			\else
			\language=\csname l@#1\endcsname
			\fi
			#2}}
	\providecommand{\BIBdecl}{\relax}
	\BIBdecl
	
	\bibitem{fast_shapelets}
	T.~Rakthanmanon and E.~Keogh, ``Fast shapelets: A scalable algorithm for
	discovering time series shapelets,'' in \emph{ICDM}, 2013, pp. 668--676.
	
	\bibitem{matrix_profile}
	C.-C.~M. Yeh, Y.~Zhu, L.~Ulanova, N.~Begum, Y.~Ding, H.~A. Dau, Z.~Zimmerman,
	D.~F. Silva, A.~Mueen, and E.~Keogh, ``Time series joins, motifs, discords
	and shapelets: A unifying view that exploits the matrix profile,''
	\emph{DMKD}, vol.~32, no.~1, pp. 83--123, Jan. 2018.
	
	\bibitem{frm94}
	C.~Faloutsos, M.~Ranganathan, and Y.~Manolopoulos, ``Fast subsequence matching
	in time-series databases,'' in \emph{SIGMOD}, 1994, pp. 419--429.
	
	\bibitem{vldb12}
	H.~Zhu, G.~Kollios, and V.~Athitsos, ``A generic framework for efficient and
	effective subsequence retrieval,'' in \emph{VLDB}, 2012, pp. 1579--1590.
	
	\bibitem{gmatch02}
	Y.-S. Moon \emph{et~al.}, ``General match: A subsequence matching method in
	time-series databases based on generalized windows,'' in \emph{SIGMOD}, 2002,
	pp. 382--393.
	
	\bibitem{tods11}
	P.~Papapetrou, V.~Athitsos, M.~Potamias, G.~Kollios, and D.~Gunopulos,
	``Embedding-based subsequence matching in time-series databases,''
	\emph{TODS}, vol.~36, no.~3, pp. 17:1--17:39, Aug. 2011.
	
	\bibitem{vldb07}
	W.-S. Han, J.~Lee, Y.-S. Moon, and H.~Jiang, ``Ranked subsequence matching in
	time-series databases,'' in \emph{VLDB}, 2007, pp. 423--434.
	
	\bibitem{kdd12}
	T.~Rakthanmanon, B.~Campana \emph{et~al.}, ``Searching and mining trillions of
	time series subsequences under dynamic time warping,'' in \emph{SIGKDD},
	2012, pp. 262--270.
	
	\bibitem{branlard2009wind}
	E.~Branlard, ``Wind energy: On the statistics of gusts and their propagation
	through a wind farm.''
	
	\bibitem{sakoe}
	H.~Sakoe and S.~Chiba, ``Dynamic programming algorithm optimization for spoken
	word recognition,'' \emph{TSP}, vol.~26, no.~1, pp. 43--49, Feb 1978.
	
	\bibitem{vldb00}
	B.-K. Yi and C.~Faloutsos, ``Fast time sequence indexing for arbitrary lp
	norms,'' in \emph{VLDB}, 2000, pp. 385--394.
	
	\bibitem{lb_paa}
	Y.~Zhu and D.~Shasha, ``Warping indexes with envelope transforms for query by
	humming,'' in \emph{SIGMOD}, 2003, pp. 181--192.
	
	\bibitem{r-tree}
	H.~Alborzi and H.~Samet, ``Execution time analysis of a top-down r-tree
	construction algorithm,'' \emph{Inf. Process. Lett.}, vol. 101, pp. 6--12,
	2007.
	
	\bibitem{hbase}
	``{Apache HBase},'' \url{http://hbase.apache.org}.
	
	\bibitem{UCRArchive}
	Y.~Chen, E.~Keogh, B.~Hu, N.~Begum, A.~Bagnall, A.~Mueen, and G.~Batista, ``The
	ucr time series classification archive,''
	\url{www.cs.ucr.edu/~eamonn/time_series_data/}.
	
	\bibitem{Fu2008}
	A.~W.-C. Fu, E.~Keogh, L.~Y.~H. Lau, C.~A. Ratanamahatana, and R.~C.-W. Wong,
	``Scaling and time warping in time series querying,'' \emph{The VLDB
		Journal}, vol.~17, no.~4, pp. 899--921, Jul 2008.
	
	\bibitem{edbt17}
	Y.~Li, B.~Tang, L.~H. U, M.~L. Yiu, and Z.~Gong, ``Fast subsequence search on
	time series data ({Poster Paper}),'' in \emph{EDBT}, 2017, pp. 514--517.
	
	\bibitem{dualmatch01}
	Y.-S. Moon, K.-Y. Whang, and W.-K. Loh, ``Duality-based subsequence matching in
	time-series databases,'' in \emph{ICDE}, 2001, pp. 263--272.
	
	\bibitem{multi07}
	S.-H. Lim, H.-J. Park, and S.-W. Kim, ``Using multiple indexes for efficient
	subsequence matching in time-series databases,'' in \emph{DASFAA}, 2006, pp.
	65--79.
	
	\bibitem{lb_keogh}
	E.~Keogh and C.~A. Ratanamahatana, ``Exact indexing of dynamic time warping,''
	\emph{KIS}, vol.~7, no.~3, pp. 358--386, Mar. 2005.
	
	\bibitem{icde18}
	R.~Neamtu, R.~Ahsan, E.~Rundensteiner, G.~N.~Sarkozy, E.~Keogh, A.~Dau,
	C.~Nguyen, and C.~Lovering, ``Generalized dynamic time warping: Unleashing
	the warping power hidden in point-wise distances,'' in \emph{ICDE}, 2018.
	
	\bibitem{adaptive}
	K.~Zoumpatianos, S.~Idreos, and T.~Palpanas, ``Indexing for interactive
	exploration of big data series,'' in \emph{SIGMOD}, 2014, pp. 1555--1566.
	
	\bibitem{vldb16}
	R.~Neamtu, R.~Ahsan, E.~Rundensteiner, and G.~Sarkozy, ``Interactive time
	series exploration powered by the marriage of similarity distances,''
	\emph{Proc. VLDB Endow.}, vol.~10, no.~3, pp. 169--180, Nov. 2016.
	
\end{thebibliography}
